\documentclass[aps,pra,twocolumn,showpacs,superscriptaddress]{revtex4-2}
\usepackage{amsmath}
\usepackage{amssymb}
\usepackage{mathtools}
\usepackage{bbold}

\usepackage{graphicx}
\usepackage{color}
\renewcommand{\vec}[1]{\mathbf{#1}}

\begin{document}
\title{Modeling Particle Loss in Open Systems \\
using Keldysh Path Integral and Second Order
Cumulant Expansion}

\author{Chen-How Huang }
 \affiliation{Donostia International Physics Center (DIPC), 20018 Donostia-San Sebastian, Spain}
 \author{Thierry Giamarchi}
 \affiliation{Department of Quantum Matter Physics, University of Geneva, 1211 Geneva, Switzerland.}
\author{Miguel A. Cazalilla }
  \affiliation{Donostia International Physics Center (DIPC), 20018 Donostia-San Sebastian, Spain}
\affiliation{Ikerbasque, Basque Foundation for Science, 48013 Bilbao, Spain}
\date{\today}
\begin{abstract}
For open quantum systems,   integration of the bath degrees of freedom
using the second order cumulant expansion in the Keldysh path integral 
provides an alternative derivation of the  effective action for  
systems coupled to general baths. The baths can be interacting and not necessarily
Markovian.  Using this method in the Markovian limit, we compute the particle loss  dynamics in various models of ultra-cold atomic gases including a one-dimensional Bose-Hubbard model with two-particle losses and a multi-component Fermi gas with interactions tuned by an optical Feshbach resonance. We explicitly demonstrate that the limit of strong two-body losses can be treated  by formulating an indirect loss scheme to describe the bath-system coupling.  The particle-loss  dynamics thus obtained is valid at all temperatures. For the one-dimensional  Bose-Hubbard model, we
compare it to solutions of the phenomenological rate equations. The latter are shown to be accurate at high temperatures.
\end{abstract}

\maketitle

\section{Introduction}

Open quantum systems constitute one of the most interesting challenges of the field, both from a fundamental point of view and of course in connection with applications or experimental realizations. This class of problems offers an interesting
interplay between coherent Hamiltonian dynamics and incoherent, dissipative dynamics emerging from  coupling to the environment~\cite{breuer2002,daley_quantum_2014, ashida_non-hermitian_2020}. From a fundamental point of view, after being  
considered as a nuisance due to the destruction of coherence, dissipative processes are now regarded as a resource allowing to engineer novel quantum states  of matter~\cite{MuellerZoller2012,CaballarWatanabe2014,SyassenDuerr2008,BarreiroBlatt2011} or 
to assist for quantum transport properties \cite{VicianiCaruso2015, MaierRoos2019,dolgirev_impurity_noise}. Among the various types of dissipative processes, particle losses have played a special role, and recent experiments in cold atomic systems have allowed to control them in an exquisite manner. This is for example the case of losses in weakly interacting 
Bose gases~\cite{BarontiniOtt2013, LabouvieOtt2016} or fermionic systems where local losses are realized~\cite{CormanEsslinger2019,LebratEsslinger2019}. More recently, cavities  have also provided  an interesting playground for this kind of 
physics~\cite{mivehvar_cavity_review,jones_photoassociation_review,konishi_polariton_cavity}.

 Theoretically  studying dissipative phenomena is a considerable challenge, and several approaches have been used in order to deal with the coupling to the environment, which is often modelled as a bath. One common approach  resorts to non-Hermitian Hamiltonians~\cite{Yamamoto_PhysRevLett.123.123601,Nakagawa2_PhysRevLett.124.147203,Yamamoto2_PhysRevB.105.205125,Ashida_AdvanceinPhys_2020}.
The resulting (non-Hermitian) models can be  analyzed using e.g.~ the powerful tools of integrability~\cite{Nakagawa_PhysRevLett.121.203001,Yamamoto3_PhysRevB.107.045110} as well as  various field-theoretic and numerical techniques~\cite{Yamamoto_PhysRevLett.123.123601,Nakagawa2_PhysRevLett.124.147203} which
include bosonization and the renormalization 
group~\cite{Yamamoto2_PhysRevB.105.205125,Yamamoto3_PhysRevB.107.045110,han2023complex}.
However, because such treatments often neglect or at most treat in an approximate
way the so-called quantum-jump term of the Lindblad master equation, it is difficult
to assess how reliably they can describe the dynamics of the system observables.
Other approaches deal with the full
Lindblad master equation~\cite{damanet_controlling_2019,damanet_reservoir_2019a,damanet_reservoir_2019b,YamamotoKawakami2021,MazzaSchiro2023} assuming the bath is Markovian~\cite{breuer2002,auletta_fortunato_parisi_2009}. In this work, we are concerned
with another generic method to deal with such out of equilibrium systems, namely 
the Keldysh formalism and its path integral formulation~\cite{kamenev_2011,Sieberer_IOP_2016}.

 Approaches based on the path integral have a long history beginning with the pioneering work of Feynman and Vernon~\cite{FEYNMAN1963118}. They have  been often used
in the context of quantum dissipation where the coupling to ohmic, subohmic or superohmic baths generates an effective long-range interaction in imaginary 
time~\cite{Sols_PhysRevB.36.7775,Cazalilla_PhysRevLett.97.076401,Lobos_PhysRevB.82.104517,Malatsetxebarria_PhysRevA.88.063630}. In a more general context, they have also been used to describe non-equilibrium dynamics in the presence of coupling to baths 
 \cite{kamenev_2011,Husmann2015},  time dependent environments 
\cite{dolgirev_impurity_noise,jin_noise_exact_1D} 
 and single particle losses 
\cite{Sieberer_IOP_2016,froeml_ultracold_2020,visuri_losses_zeno_long,huang_MAR_cold}. 

Despite these important developments, several phenomena linked to the coupling to the environment still remain elusive. In particular, in ultracold atomic systems the  dynamics in the presence particle losses involve \emph{a priori}  processes beyond single-particle losses and include  two- or three-particle losses as well. 
Quite generally, the description of particle losses has assumed the validity of  phenomenological rate equations of the form:
\begin{equation}\label{eq:phenon}
    \frac{d n }{dt}=-\gamma_1 n- \gamma_2 n^2
\end{equation}
where  $n$ is the particle density and $\gamma_1$ and $\gamma_2$ are the one- and two-particle loss rates, respectively. However, the correct functional form and range of validity  of these phenomenological equations and how to account for many-particle effects are  still not  fully understood.  For ultra-cold atoms near Feshbach resonances, a theoretical description of the two-body loss rate, $\gamma_2$, was developed based on S-matrix calculation in Ref.~\cite{Bohn_PhysRevA.56.1486} and some properties of three-body losses could be determined from Bethe ansatz solutions~\cite{cheianov_threebody_losses}.
Braaten et al.~\cite{Braaten_2017}  combined effective field theory 
with the Lindblad master equation to obtain a universal relation for the two-atom inelastic loss rate for  two-species Fermi gases.

  In this work, in order to go beyond the phenomenological description of particle losses provided by Eq.~\eqref{eq:phenon} and be able to account for many-particle effects,  we develop a path integral formalism that relies on the second order cumulant expansion. The formalism can  in principle also deal with interacting as well as non-Markovian baths. However, here we demonstrate it by studying the Markovian dynamics of many-particle systems in the presence of  one- and two-particle losses and briefly discuss how to go beyond the Markovian limit in the Appendix~\ref{app:corr}. In the absence of interactions in the bath, the resulting approach is exact for one-particle losses and approximately valid for two-particle loss problems at weak coupling.  In order to deal with the strong coupling regime, we also introduce an indirect two-body loss scheme. We discuss two experimentally relevant examples of this indirect-loss scheme (a lossy one-dimensional  Bose-Hubbard model and a multi-component Fermi gas near an optical Feshbach resonance) and  explicitly derive the loss rate equations. In both cases, we find their functional form deviates from Eq.~\eqref{eq:phenon}.  In the  case of the lossy Bose-Hubbard model, we explicitly show that, at low-temperatures where quantum coherence is important, predictions of the microscopic theory  deviate  substantially from those obtained using the phenomenological equation~\eqref{eq:phenon}.

The rest of this article is organized as follows. In Sec.~\ref{sec:formalism} we
describe the derivation of the Keldysh path integral using the 2nd order cumulant expansion. We first illustrate  
the method with a model consisting of a single-mode (fermonic or bosonic) oscillator coupled to a bath to which particles can be lost (cf. Sec.~\ref{sec:singlemode}). The general formalism is discussed 
in Sec.~\ref{sec:general}. However, since only the results of subsection~\ref{sec:singlemode}
will be used in the rest of the article, Sec.~\ref{sec:general} can be skipped 
on a first reading. Sec.~\ref{sec:onebody} briefly describes how the results 
of Sec.~\ref{sec:singlemode} are applied to describe a quantum gas with one-body losses. Sections~\ref{sec:lossybhm} 
and \ref{sec:ofr} deal with the applications of the formalism to two 
 models for relevant for the physics of ultracold atomic gases with two-body losses:
 Sec.~\ref{sec:lossybhm}
is concerned with a lossy Bose-Hubbard model in one dimension, whereas Sec.~\ref{sec:ofr} deals with the model of a multi-component Fermi gas near  optical Feshbach resonance.  Some technical details  and useful results are described to the Appendices.

\section{Formalism}\label{sec:formalism}

\subsection{Single-mode case}\label{sec:singlemode}

To begin with, let us illustrate the method by considering a single bosonic or 
fermionic mode (system $A$) coupled  to a bath ($B$) by means of a quadratic Hamiltonian:
\begin{align}
H_{A} &= \epsilon_0 a^{\dag} a + H_{A,\mathrm{int}},\\
H_{B} &= \sum_{\alpha}\omega_{\alpha} b^{\dag}_{\alpha} b_{\alpha} + H_{B,\mathrm{int}},\\
H_{AB}(t) &=  f(t) \sum_{\alpha}  \left[ g_{\alpha} a^{\dag} b_{\alpha} 
+ g^{*}_{\alpha} b^{\dag}_{\alpha} a \right]
\end{align}
Here $a,a^{\dag}$ describe a bosonic (fermionic) mode in the $A$ and $b_{\alpha},b^{\dag}_{\alpha}$
a set of bosonic (fermionic) modes in  $B$ labelled by a continuum index $\alpha$.
We shall assume that $A$ and bath $B$ are in equilibrium (not necessarily with
each other) at $t=-\infty$ and the interaction between them $H_{AB}(t)$ is switched according
to a protocol described by the function $f(t)$. Both $A$ and $B$ can be interacting systems with
\emph{e.g.}  $H_{B,\mathrm{int}} = g_{\rm{int}}\sum_{\alpha\beta\gamma\delta} b^{\dag}_{\alpha}b^{\dag}_{\beta}b_{\gamma}b_{\delta}$.

The Keldysh generating functional~\cite{kamenev_2011} for the system introduced above can be written as follows:
\begin{equation}
Z[\bar{V},V] = \int D[\bar{a}a] D[\bar{b}_{\alpha} b_{\alpha}] \, e^{i S}.
\end{equation}
The Keldysh action  is $S = S_A
+ S_B + S_V +  S_{AB}$, where
\begin{align}
S_A &= \int_C dt \, \left[ i\bar{a} \partial_t  a - H_{A}\right],\\ 
S_{B}&= \int_C dt\: \left[ i \sum_{\alpha} \bar{b}_{\alpha} \partial_t b_{\alpha} - H_{\mathrm{int},B}\right],\\ 
S_V &= \int_C dt \,\left[ \bar{a}(t) V(t) + \bar{V}(t) a(t) \right],\\
S_{AB} &= - \sum_{\alpha} \int_C dt  \: f(t) \left[ g_{\alpha} \bar{a}(t) b_{\alpha}(t) \right. \notag\\
 &\qquad\qquad\left. + g^*_{\alpha} \bar{b}_{\alpha}(t) a(t) \right].
\end{align}
In the above expressions, $C$ is the Keldysh contour, which runs from $t=-\infty$ to $t=+\infty$
and back to $t=-\infty$~\cite{kamenev_2011}; $\bar{V},V$ are sources that couple the system 
degrees of freedom: Any $n$-point correlation of the system $A$ can be obtained by conveniently
taking functional derivatives of $Z[\bar{V},V]$ with respect to $\bar{V}$ and $V$.
The generating functional is normalized so that $Z[\bar{V}=0,V=0] = 1$ since in the absence of external sources it merely describes the unitary evolution of the initial state from $t=-\infty$ to $+\infty$ and back to $t=-\infty$. 

Starting from the above functional integral representation, we can define the 
Feynman-Vernon influence functional~\cite{FEYNMAN1963118}  $\mathcal{F}[\bar{a},a]$  as the result of formally integrating  out the bath degrees
of freedom, i.e.
\begin{equation}
\mathcal{F}[\bar{a},a]=\int \, D[\bar{b}_{\alpha} b_{\alpha}]\, e^{i \left( S_B +S_{AB}\right)}
\end{equation}
It is often not possible to obtain $\mathcal{F}$ in a closed form and 
therefore we have to resort to approximations. For a general (e.g. interacting) bath that
is weakly coupled to the system  the 2nd order cumulant expansion provides
a good starting point to capture the dissipative dynamics induced by the bath.
Furthermore, it is exact if the bath Hamiltonian $H_B$ and the coupling
$H_{AB}$ are quadratic and linear in the fields 
$\bar{b}_{\alpha},b_{\alpha}$, respectively. Using the cumulant expansion to
second order (see Appendix~\ref{app:cumu}), we  approximate:
\begin{align}
\mathcal{F}[\bar{a},a] &= \langle e^{i S_{AB}} \rangle_B \notag\\
&\simeq  \exp\left[
 i \langle S_{AB}\rangle_B 
 - \frac{1}{2} \left( \langle S^2_{AB} \rangle_{B} - \langle S_{AB} \rangle_{B}^2 \right) 
\right]\notag\\ 
&= e^{i \mathcal{L}[\bar{a},a]},
\end{align}
where $\langle\cdots\rangle_B$ stands for average over the bath degrees of freedom.
Taking $\langle S_{AB}\rangle_B = 0$ after assuming that the bath conserves the number of bath excitations in its initial state, we obtain the following dissipative effective  action:
\begin{multline}
\mathcal{L}= \frac{i}{2} \sum_{\alpha}\int_C dt_1 dt_2 \, |g_{\alpha}|^2 f(t_1) f(t_2)\\ 
\times \left[ \langle b_{\alpha}(t_1) \bar{b}_{\alpha}(t_2) \rangle_B\:  \bar{a}(t_1) a(t_2) +\right.\\
\qquad\qquad \left. +  \langle \bar{b}_{\alpha}(t_1) b_{\alpha}(t_2) \rangle_B \: a(t_1) \bar{a}(t_2) \right] \\
= \frac{i}{2} \sum_{m,n=\pm}\int dt_1 dt_2   v^{mn}(t_1, t_2) \bar{a}_{m}(t_1) a_{n}(t_2).
\end{multline}
In the last expression, we have split the integrals over the Keldysh contour $C$. After rewriting them as a single integral where $t_1,t_2$ run from $-\infty$ to $+\infty$, we have introduced the subindices $m,n=\pm$ to denote on which branch
of $C$ the time argument of the fields $a, b$, etc lies. We have also 
introduced the functions $v^{mn}(t_1,t_2) = f(t_1) f(t_2) g^{mn}(t_1-t_2)$, where
\begin{align}\label{eq:bc}
g^{mn}(t_1-t_2) = s_{n} s_{m} \sum_{\alpha} |g_{\alpha}|^2 \left[ \langle  b_{\alpha m}(t_1) \bar{b}_{\alpha n}(t_2) \rangle_B \right. \notag \\ 
\left. + z \langle \bar{b}_{\alpha n}(t_2) b_{\alpha m}(t_1)\rangle_B \right].
\end{align}
In the above expression $s_{m = \pm} = \pm 1$ and $z = +1$ ($z=-1$)  for bosons (fermions). 
  In the Appendix, within Markovian limit, we show that  $g^{mn}(t_1-t_2)\sim\delta(t_1-t_2)$. Thus, we arrive at the following expression for the effective couplings (see Appendix~\ref{app:corr} for more detail),
\begin{align}
v^{++}(t_1,t_2)  &\simeq   \nu_0  |\langle g\rangle f(t_1) |^2   \delta(t_1-t_2) , \\
v^{--}(t_1,t_2)  &\simeq  \nu_0  |\langle g\rangle f(t_1) |^2   \delta(t_1-t_2),\\
v^{-+}(t_1,t_2)&\simeq -2 \nu_0 |\langle g\rangle f(t_1) |^2   \delta(t_1-t_2)\\
v^{+-}(t_1,t_2) &\simeq 0,
\end{align}
where $\langle g\rangle$ is an average system-bath coupling constant. Upon denoting $\gamma(t) = \nu_0 |\langle g\rangle f(t)|^2$ for the loss rate of particles to the bath, we obtain the following result:
\begin{multline}\label{eq:single mode}
\mathcal{L} = -i \int dt  \: \gamma(t) \left[    \bar{a}_{-}(t) a_{+}(t) \right. \\
\left. -\tfrac{1}{2} \left(  \bar{a}_{+}(t) a_{+}(t) + \bar{a}_{-}(t)  a_{-}(t) \right) \right],
\end{multline}
This is the dissipative part of the action characteristic of a Markovian bath. It
 can also be obtained from the evolution of the  density matrix
according to the Lindblad master equation (see Appendix~\ref{app:recap}, which is based 
on Ref.~\cite{Sieberer_IOP_2016} and references therein). 

Finally, let us compute the particle loss in the system $A$ caused by switching on the
coupling to the bath $B$ at $t=0$ for an infinitesimal time $\delta t$. This calculation
can be carried out by the path integral version of time-dependent perturbation theory, i.e.
by perturbatively expanding the effective dissipative  action to leading order in $\mathcal{L}$,
which yields:

\begin{align}
\langle a^{\dag}(t) a(t) \rangle &= \int D[\bar{a}a] \, \bar{a}_{-}(t) a_{+}(t) \:
e^{i \left( S_A + \mathcal{L} \right) }  ,\notag \\
& = \int D[\bar{a}a] \, \bar{a}_{-}(t) a_{+}(t) \left [1 +  \mathcal{L} +  O( \mathcal{L}^2) \right]  \:
e^{i  S_A } ,\notag\\
&\simeq \langle \bar{a}_{-}(t) a_{+}(t)\rangle_A + i \langle \bar{a}_{-}(t) a_{+}(t) \mathcal{L} \rangle_A, \\
&= n_a  +  \gamma \int^{+\infty}_0 dt^{\prime} \langle \bar{a}_{-}(t) a_{+}(t)  \bar{a}_{-} (t')a_{+}(t')\rangle_A \notag\\
&\, - \tfrac{1}{2} \gamma \int^{+\infty}_0 dt^{\prime}  \langle \bar{a}_{-}(t) a_{+}(t)  \bar{a}_{+} (t')a_{+}(t')\rangle_A \notag\\
&\, - \tfrac{1}{2} \gamma \int^{+\infty}_0 dt^{\prime}  \langle \bar{a}_{-}(t) a_{+}(t)  \bar{a}_{-} (t')a_{-}(t')\rangle_A,
\end{align}
where we have set $\gamma(t) =\gamma \theta(t)$ and denoted 
$n_a = \langle a^{\dag} a\rangle $, which is the occupation in the initial state (i.e. for $t < 0$).
Assuming the system $A$ is non-interacting, we have
\begin{align}
 \langle \bar{a}_{-}(t) a_{+}(t) \bar{a}_{-}(t^{\prime}) a_{+}(t^{\prime})  \rangle_A &=   z n^2_a,\\
\langle \bar{a}_{-}(t) a_{+}(t) \bar{a}_{+}(t^{\prime}) a_{+}(t^{\prime}) \rangle_A &= 
\tilde{\theta}(t-t^{\prime}) n_a(1+z n_a) \notag\\
&\qquad + \theta(t^{\prime}-t) z n^2_a,\\
\langle \bar{a}_{-}(t) a_{+}(t) \bar{a}_{-}(t^{\prime}) a_{-}(t^{\prime}) \rangle_A &= \theta(t-t^{\prime}) n_a(1+z n_a) \notag \\ &\quad + \tilde{\theta}(t^{\prime} - t) zn^2_a.
\end{align}
Hence,
\begin{equation}
   n_a(t) =  \langle a^{\dag}(t) a(t) \rangle =n_a - \gamma n_a \int^{t}_0  dt^{\prime} \, \theta(t-t^{\prime}). 
\end{equation}
Setting $t=\delta t\gg D^{-1}$ (where $D^{-1}$ is the characteristic response time of the bath, see Appendix~\ref{app:corr}), we arrive at the following one-body loss rate equation:
\begin{equation}
\frac{ d n_a(t)}{dt} = -\gamma n_a(t),
\end{equation}
In the right hand-side, with accuracy $O(\gamma \delta t)$, we have replaced $n_a$  by $n_a(\delta t)$.

\subsection{General case} \label{sec:general}

 In this section, we generalize the above results. Since only the results obtained in previous section will be necessary the applications discussed in this work, this section can be entirely skipped on a first reading.  The starting point is again the Hamiltonian describing the unitary dynamics of  a system ($A$) coupled to a bath ($B$).
\begin{equation}
H=H_A(t)+H_B+H_{AB}(t),\label{eq:ham}
\end{equation}
where $H_A(t)$, $H_B$, and $H_{AB}(t)$ denote the  Hamiltonians of the system, bath and their coupling, respectively. We have assumed that, in general, the system Hamiltonian and its
coupling to the bath can be explicitly time-dependent. We shall consider a rather general form of the coupling between system and bath:
\begin{align}
H_{AB}(t) &=\sum_{\vec{q},\vec{p}} \left[  g_{\vec{q}\vec{p}}(t) A^{\dag}_{\vec{q}}
B_{\vec{p}}+ \mathrm{h.c.} \right],\label{eq:sysbath}
\end{align}
where 
\begin{align}
A_{\vec{q}}= a^{\dag}_{\bar{q}_1}\cdots a^{\dag}_{\bar{q}_N} a_{q_1}a_{q_2}\cdots a_{q_{N^{\prime}}},\\ 
B_{\vec{p}}=b^{\dag}_{\bar{p}_1}\cdots b^{\dag}_{\bar{p}_M}  b_{p_1}b_{p_2}\cdots b_{p_{M^{\prime}}}
\end{align}
are  products of  arbitrary number operators (with $N\neq N^{\prime}$ and $M\neq M^{\prime}$ in general) acting  either on the system or the bath;   $\vec{p} =\{ p_1,\cdots,p_{M^{\prime}}; \bar{p}_1,\cdots,\bar{p}_{M}\}$ and $\vec{q} = \{ q_1,\cdots,q_{N};\bar{q}_1,\cdots,\bar{q}_{N^{\prime}}\}$ are the quantum numbers carried by those operators;  $g_{\vec{ qp}}(t)$ are the set of system-bath couplings. The system-bath coupling is switched on according to a certain protocol that determines the explicit 
time dependence of the $g_{\vec{qp}}(t)$.

%
%
%
%

In general, we are not able to obtain the Feynman-Vernon influence functional $\mathcal{F}$ exactly and here we will resort to a second order cumulant 
expansion:
\begin{align}
\mathcal{F}[\bar{a},a] &= \exp\left[ i\langle S_{\text{AB}}\rangle_B- \tfrac{1}{2} (\langle S_{\text{AB}}^2\rangle_B-\langle S_{\text{AB}} \rangle^2_B )+\cdots \right]\notag\\
&= e^{i\mathcal{L}[\bar{a},a]}.
\end{align}
Explicitly, for the system-bath coupling introduced in Eq.~\eqref{eq:sysbath} the first order correction takes the  form:
\begin{align}
\langle S_{AB} \rangle_B &=  \sum_{\vec{p}\vec{q}}  
\int_C dt \left[ g_{\vec{q p}}(t) \bar{A}_{\vec{q}}(t) \langle  B_{\vec{p}}(t)\rangle_B \right.  \notag\\
& \left. \qquad\qquad + g^*_{\vec{q p}}(t)  \langle \bar{B}_{\vec{p}}(t)\rangle_B  A_{\vec{q}}(t)  \right]
\end{align}
Note that this  term does not describe dissipation and only modifies the unitary evolution of the system $A$. Thus, it can be conveniently absorbed into $H_S$ by defining the operators $B_{\vec{p}}$ and $B^{\dag}_{\vec{p}}$ in Eq.~\eqref{eq:sysbath} to have zero 
averages, i.e. $\langle B_{\vec{p}}\rangle_B = \langle B^{\dag}_{\vec{p}}\rangle_B = 0$. In addition, this is automatically fulfilled if $B$ and $B^{\dag}$ change the number of 
particle/excitations in the bath but the bath Hamiltonian $H_B$ and its initial state conserve this number. Therefore, in what follows, we set $\langle S_{\text{AB}} \rangle_B  = 0$ and do not discuss it any further.

The second order correction can be brought to the following form in terms
of  bath correlators:
\begin{widetext}
\begin{align}
\mathcal{L} = \tfrac{i}{2!}\langle S_{AB}^2\rangle_B   &= \frac{i}{2}  \sum_{\vec{ q_2 q_1}}  \int_C dt_1dt_2   \biggl[
u_{\vec{q_1}\vec{q_2}}(t_1,t_2)\bar{A}_{\vec{q_1}}(t_1)\bar{A}_{\vec{q_2}}(t_2) + \bar{u}_{\vec{q_1}\vec{q_2}}(t_1,t_2)  A_{\vec{q_1}}(t_1)A_{\vec{q_2}}(t_2)  \notag\\
 & \qquad  \qquad  + v_{\vec{q_1}\vec{q_2}}(t_1,t_2)  \bar{A}_{\vec{q_1}}(t_1)A_{\vec{q_2}}(t_2) 
  +  \bar{v}_{\vec{q_1}\vec{q_2}}(t_1,t_2) A_{\vec{q_1}}(t_1)\bar{A}_{\vec{q_2}}(t_2)  \biggr],
 \label{eq:eff}
\end{align}
where
\begin{align}
u_{\vec{q}_1,\vec{q}_2}(t_1,t_2)&=\sum_{\vec{p}_1\vec{p}_2}g^*_{\vec{p_1}\vec{q_1}}(t_1) g^*_{\vec{p_2}\vec{q_2}}(t_2)
C_{BB}(\vec{p}_1 t_1,\vec{p}_2 t_2),\\
\bar{u}_{\vec{q_1}\vec{q_2}}(t_1,t_2) &= 
\sum_{\vec{p_1p_2}} g_{\vec{p_1}\vec{q_1}}(t_1) g_{\vec{p_2}\vec{q_2}}(t_2) C_{\bar{B} \bar{B}}(\vec{p}_1 t_1, \vec{p}_2 t_2),\\
v_{\vec{q}_1,\vec{q}_2}(t_1,t_2)&=\sum_{\vec{p}_1\vec{p}_2}g^*_{\vec{p_1}\vec{q_1}}(t_1) g_{\vec{p_2}\vec{q_2}}(t_2)
C_{B \bar{B}}(\vec{p}_1 t_1,\vec{p}_2 t_2),\\
\bar{v}_{\vec{q_1}\vec{q_2}}(t_1,t_2) &= 
\sum_{\vec{p_1p_2}} g_{\vec{p_1}\vec{q_1}}(t_1) g^*_{\vec{p_2}\vec{q_2}}(t_2) C_{\bar{B}B}(\vec{p}_1 t_1, \vec{p}_2 t_2),
\label{eq:couplings}
\end{align}
\end{widetext}
In the above equation, the following notation has been introduced ($X,Y = B, \bar{B}$):
\begin{equation}
C_{XY}(\vec{p}_1 t_1,\vec{p}_2 t_2) = 
\langle  T_C\left[ X(\vec{p_1},t_1)Y(\vec{p_2},t_2) \right]\rangle_B
\end{equation}
for the bath two-point correlation functions of operators $B,B^{\dag}$.
 In the context of the Keldysh path integral each one of the above correlation functions becomes  a $2\times 2 $ matrix in the superindices $m,n = \pm$ after expanding the integrals over $C$ so that $t_1,t_2$ run from $-\infty$ to $+\infty$. The superindices are inherited by the couplings $u,v,\bar{u},\bar{v}$ introduced in Eq.~\eqref{eq:couplings}, which also become the $2\times 2$ matrices $u^{mn}_{\vec{q}_1 \vec{q}_2}(t_1,t_2)$, $v^{mn}_{\vec{q}_1 \vec{q}_2}(t_1,t_2)$,
 etc.
 
Explicit consideration of the time-dependence of the coupling matrices $u^{mn}_{\vec{q}_1 \vec{q}_2}(t_1,t_2),v^{mn}_{\vec{q}_1 \vec{q}_2}(t_1,t_2)$, etc, for a given bath may allow to identify  regimes where the Markovian approximation applies. This is typically the case when the response of the bath is much faster than the characteristic time scale of the system dynamics. Thus, in the Markovian regime we can assume that 
$u^{mn}_{\vec{q}_1 \vec{q}_2}(t_1,t_2) \propto \delta(t_1-t_2)$, etc (although some of them may also vanish as it was the case in the previous example).   This provides an additional simplification of the  second order term of the cumulant expansion and leads to the following result:
\begin{equation}
\mathcal{L} =  \mathcal{L}_{N} + \mathcal{L}_{A},
\end{equation}
where $S_N$ is the ``normal'' part: 
\begin{align}
\mathcal{L}_N &= \frac{i}{2} \sum_{\vec{ q_2 q_1},m,n=\pm} \int dt \,   
 \tilde{v}^{mn}_{\vec{q_1}\vec{q_2}}(t) \bar{A}_{\vec{q_1},m}(t) A_{\vec{q_2},n}(t)
\label{eq:S2eff}
\end{align}
and $\mathcal{L}_A$ is the ``anomalous'' part 
\begin{align}
\mathcal{L}_A &=  \frac{i}{2}\sum_{\vec{ q_2 q_1},m,n=\pm} \int dt \,  \biggl[  \bar{u}^{mn}_{\vec{q_1}\vec{q_2}}(t) A_{\vec{q_1}m}(t)A_{\vec{q_2}n}(t),\notag\\&\qquad\qquad\qquad+u^{mn}_{\vec{q_1}\vec{q_2}}(t) \bar{A}_{\vec{q_1}m}(t)\bar{A}_{\vec{q_2}n}(t)  \biggr]
\label{eq:S2eff2}
\end{align}
In the above expressions, we have introduced the following system-bath coupling matrices:
%

\begin{equation}
\tilde{v}^{mn}_{\vec{q_1}\vec{q_2}}(t) = s_m s_n \left[ v^{mn}_{\vec{q_1}\vec{q_2}}(t) + z \bar{v}^{nm}_{\vec{q_2}\vec{q_1}}(t)  \right],\\
\end{equation}
%
where $z=-1$ ($z=+1$) if the operator products $A_{\vec{q}},A^\dag_{\vec{q}}$ in $H_\text{SB}$ have fermionic (bosonic) statistics, and $s_{+} = +1 \, (s_{-}=-1)$ . We can classify the different terms according to the type of time arguments of the system  degrees of freedom $A_{\vec{q}}(t),\bar{A}_{\vec{q}}(t)$. The first two terms on the right-hand site contain  $A_{\vec{q}m}(t),\bar{A}_{\vec{q}n}(t)$ whose time arguments lie on different branches of the  Keldysh contour $C$ and therefore $m\neq n$. These terms correspond to the so-called quantum jump terms of the Lindblad master equation (see Sec.~\ref{sec:onebody} and Appendix~\ref{app:recap}). The remaining terms contain  $A_{\vec{q}m}(t),\bar{A}_{\vec{q}n}(t)$ with time arguments lying on the same branch of $C$, i.e. with $m=n$. Such terms contribute to the anti-commutator terms of the Lindblad master equation in the operator language (see Appendix~\ref{app:recap}) and give rise to the anti-Hermitian part of the effective non-hermitian Hamiltonian in the operator language~\cite{breuer2002}.

Finally, although the last few expressions above have been derived under the assumptions of Markovianity of the bath,  we want to emphasize that the approach used here is not limited to the  Markovian regime and it can be used as a starting point to include effects beyond Markovianity. The latter are outside the range of applicability of the Lindblad master equation or its path integral formulation as introduced in Ref.~\cite{Sieberer_IOP_2016}. In the following sections, we shall consider a number of  applications to particle loss and show that, although derived using the cumulant expansion up to second order which may appear to be only valid for a weak system-bath coupling, it is  possible reformulate the models to describe the limit of very strong two-body losses exactly.

\section{One-body Loss}\label{sec:onebody}

  Before considering systems with two-body losses,  it is interesting to generalize the results of Sec.~\ref{sec:singlemode} to describe an ultracold gas coupled to a bath to which it can 
  loose one particle at a time. This section 
 largely relies on the results obtained in Sec.~\ref{sec:singlemode}. We generalize the Hamiltonian
 to a uniform gas and therefore the fields carry a momentum index $\vec{k}$. The Hamiltonian reads:

 \begin{align}
H_{A}&=\sum_{\vec{k}}\epsilon_\vec{k} a^{\dag}_\vec{k} a_\vec{k}, \\
H_{B}&=\sum_{\vec{k},\alpha} \epsilon_{\vec{k},\alpha} b^{\dag}_{\vec{k}\alpha} b_{\vec{k}\alpha}\\
H_{AB}&=  f(t) \sum_{\vec{k},\alpha}  \left[ g_{\alpha} a^{\dag}_{\vec{k}} b_{\vec{k}\alpha}+ g^*_{\alpha}\: b^\dag_{\vec{k}\alpha} a_\vec{k}\right]
\end{align}
where the fields $a_{\vec{k}}$ and $b_{\vec{k},\alpha}$ are either fermionic or bosonic.
In ultra-cold atomic gases, 
the couplings $g_{\alpha},g_{\alpha}$ are often not known from first
principles. Instead, what is measured is the loss rate of particles.  Following
the same steps as in Sect.~\ref{sec:singlemode} while keeping track the momentum
index $\vec{k}$, we arrive at the following effective  action in the Markovian limit 
($m,n=\pm$):
\begin{align}
&\mathcal{S}_{A,\mathrm{eff}} = S_A + \mathcal{L},\\
&S_A = \sum_{\vec{k},mn}\int dt\, \left[ \sigma^{3}_{mn} \bar{a}_{\vec{k}m}\left( i \partial_{t}-\epsilon_{\vec{k}} \right) a_{\vec{k}n} \right],\\
&\mathcal{L} =  -i \sum_{\vec{k},mn} \int dt\, \gamma(t) \left[ \left( \sigma^{-}_{mn} -\tfrac{1}{2}\sigma^{0}_{mn}\right) \bar{a}_{\vec{k}m} a_{\vec{k}n} \right], 
\end{align}
where we have introduced the following short-hand notations: 
$\sigma^3_{++} =  - \sigma^3_{--} = 1$,
$\sigma^{-}_{-+} = 1$,   $\sigma^0_{++} =  \sigma^{0}_{--} = 1$,
and zero otherwise.

Like in Sect.~\ref{sec:singlemode}, we can  obtain the rate of particle loss by assuming
the coupling to the bath is switched on at $t= 0$ (i.e. $f(t) = \theta(t)$) and computing
the change in the total particle number density using perturbation theory in $\cal{L}$:
\begin{align}
n_A(t) &=  \frac{1}{\Omega}\sum_{\vec{k}} \langle a^{\dag}_{\vec{k}}(t) a_{\vec{k}}(t) \rangle \\\
&= \frac{1}{\Omega}\sum_{\vec{k}} 
\int D[\bar{a}a] \, \bar{a}_{\vec{k}-}(t) a_{\vec{k}+}(t) \, e^{i\left(S_A  + \mathcal{L}\right)}
\end{align}
Following the same steps as in Sec.~\ref{sec:singlemode}, we compute the leading order change of $n_{\vec{k}}(t) = \langle a^{\dag}_{\vec{k}}a_{\vec{k}}\rangle$. Thus, we obtain rate equations  for the momentum distribution $d n_{\vec{k}}(t)/dt = -\gamma n_{\vec{k}}(t)$, and hence the rate of change of the  particle density follows:
\begin{equation}
\frac{d n_A(t)}{dt} = -\gamma n_A(t).
\end{equation}
Note that the rate of change is proportional to the density, which is characteristic of the
one-body loss process.

\section{Lossy 1D Bose Hubbard Model}~\label{sec:lossybhm}

\begin{figure}[b]
\includegraphics[width=1\columnwidth]{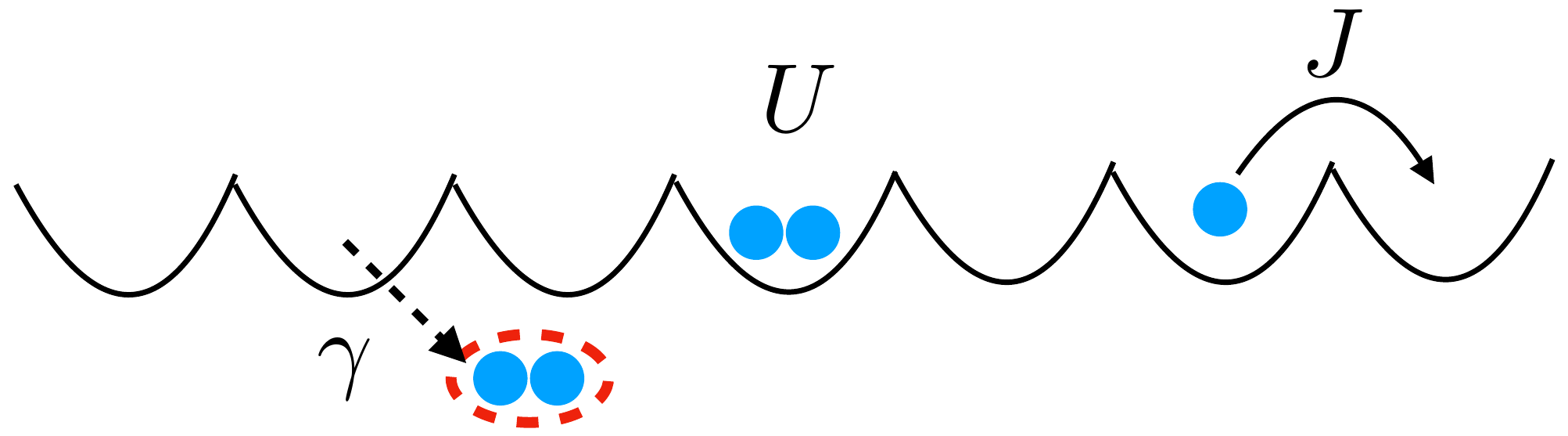}
\center
\caption{Scheme of the lossy 1D Bose-Hubbard model. $J$ is the nearest-neighbor hopping and $U$ is the onsite interaction. The loss is parametrized by  $\gamma$ is the one-body loss rate of the doublons, i.e. doubly occupied sites.} \label{fig:lossyBH}
\end{figure}
  Next, we discuss how to describe two-body losses by
coupling a system to a bath. We first consider an ultracold gas in a deep optical lattice 
which can be described by the one-dimensional Bose-Hubbard model~\cite{RevModPhys.83.1405}. 
 A  laser is applied that photo-associates   atoms 
in doubly occupied lattice sites (doublons) into molecules.  
The molecules are quickly lost from the trap, resulting in a
loss of two bosons (see Fig.~\ref{fig:lossyBH}). In the limit where the loss of the photoassociated molecule is very fast, 
the double occupancy is strongly suppressed. In other words,
in the presence of this coupling,  
the states containing doublons are rapidly projected out. However, virtual transitions
to states containing doublons can still have an effect on the system dynamics. This kind of loss dynamics was experimentally studied in~\cite{Tomita_PhysRevA_2019,Syassen_science_2008} and  theoretically described  using an approach based on an effective master equation derived in~\cite{Garcia-Ripoll_2009}, which yields a phenomenological rate equation like~\eqref{eq:phenon}. In the following, we provide a microscopic theory for the loss dynamics and compare it to the phenomenological loss equation. 

A convenient
way to describe the dynamics of the Bose-Hubbard model in a subspace containing no doublons
relies on the Jordan-Wigner transformation~\cite{RevModPhys.83.1405} that relates hard-core bosons to fermions:
\begin{equation}
c_{j} = K_j a_j, \qquad K_i = \prod_{l<j} (1-2 n_l)
\end{equation}
where $n_j = a^{\dag}_j a_j = c^{\dag}_{j}c_{j} = 0, 1$ measures the occupation of site $j$.
The transformation holds true provided $(a^{\dag}_j)^2 = a^{2}_j = 0$ (i.e. no doublons) in the
relevant Fock subspace. In this subspace, the kinetic energy of the hardcore bosons can be 
written in terms of the Jordan-Wigner fermions (see Fig.~\ref{fig:lossyBH}):
\begin{equation}
H_c = - J \sum_{j} \left[c^{\dag}_j c_{j+1} + \mathrm{h.c.} \right]. 
\end{equation}
In addition, we note that the original hopping operator of bosons $-J \sum_{i} 
\left[ a^{\dag}_{j} a_{j+1} + \mathrm{h.c.}\right]$ also allows for  transitions that create 
virtual doublons. In order to allow for
such processes, we introduce a doublon field on each site $d^{\dag}_i$ that is coupled to the
(hardcore) bosons by means of $-J\sum_{j} \left(d_{j} + d_{j+1}\right) a^{\dag}_{j} a^{\dag}_{j+1} + \mathrm{h.c.}$.  Following the Jordan-Wigner transformation, this coupling becomes:
\begin{equation}
H_{cd} = -J \sum_{j} \left[  \left(c^{\dag}_{j} c^{\dag}_{j+1} +  
c^{\dag}_{j-1} c^{\dag}_{j}\right) d_{j}   + \mathrm{h.c.} \right]. 
\end{equation}
Finally, we note that the doublon has an excitation energy equal to $U$, and therefore,
\begin{equation}
H_{d} = U \sum_{j} d^{\dag}_j d_j.
\end{equation}
This rather heuristic derivation of the Hamiltonian in the limit where the doublons are suppressed 
is confirmed below by showing that it is a convenient Hubbard-Stranovich decoupling of the effective interaction generated in the strongly interacting limit of the one-dimensional Bose-Hubbard model i.e. for  $U \gg J$~\cite{Cazalilla_PhysRevA.67.053606}.

The generating functional for the above model is:
\begin{equation}\label{eq:zvv}
Z[\bar{V},V] = \int  D[ \bar{d} d ] D[ \bar{c} c] \, e^{i S},
\end{equation}
where $S = S_c + S_d + S_{dc} + S_V$
\begin{widetext}
\begin{align}
S_c &=   \sum_{j} \int_C dt \: \left[ \bar{c}_j (i\partial_t -\mu) c_{j}  + J \left(\bar{c}_j c_{j+1} + \bar{c}_{j+1} c_j \right) \right], \label{eq:sc} \\
S_{dc} &= J\sum_j \int_C dt \left[ \left( \bar{c}_{j}\bar{c}_{j+1} + \bar{c}_{j-1}\bar{c}_{j}\right) d_j  + \bar{d}_j  \left( c_{j+1} c_j + c_{j} c_{j-1} \right) \right],\\
S_V &=\sum_j \int_C dt\, \left[ \bar{V}_j c_j + \bar{c}_j V_j\right].
 \end{align}
\end{widetext}
 As mentioned above, the doublon field will be treated as Hubbard-Stranotovich field
and therefore its ``action'' does not contain a time-derivative term $i\bar{d}\partial_t d$:
\begin{equation}
S_{d} =  - U \sum_{j} \int_C dt\, \bar{d}_j  d_j.
\end{equation}
The calculations to be described below can be also carried out including such derivative 
term, which is not important in the limit where $U\gg J$ (see next section for a  discussion where a similar situation is encountered).  We further assume that the system is coupled to a bath that removes the doublons. This coupling is described by the following term in the Keldysh action:
\begin{equation}
S_{dB} = - \sum_{j,\alpha} \int_C dt f(t) \: \left[ g_{\alpha} \bar{d}_j b_{j\alpha} + g^*_{\alpha} \bar{b}_{j\alpha} d_j \right],
\end{equation}
where the bath modes have the following quadratic action:
\begin{equation}
S_{B}= \sum_{j,\alpha}\int_C dt \, \bar{b}_{j\alpha}\left( i\partial_t - \omega_{\alpha}
 \right)b_{j\alpha}. 
 \end{equation}
We integrate out the bath following the same steps as in Sect.~\ref{sec:singlemode}, which
in the Markovian limit yields:
\begin{equation}\label{eq:BH1}
\mathcal{L}_d =  -i \sum_{j,mn}\int dt\: \gamma(t)  \left(\sigma^{-}_{mn} - \tfrac{1}{2} \sigma^0_{mn} \right)  \bar{d}_{jm} d_{jn}.
\end{equation}
Combining this term with the action for the doublon, the following effective action is obtained:
\begin{align}
S_{d,\mathrm{eff}} &= \sum_{j,mn} \int dt\, \bar{d}_{jm} G^{-1}_{mn}(t) d_{jn},\\
G^{-1}_{mn}(t) &=  -U \sigma^{3}_{mn} -i\gamma(t) \left(\sigma^{-}_{mn} - \tfrac{1}{2} \sigma^0_{mn} \right).
\end{align}
Finally, we integrate out the doublon field by making the following change of integration variables
in the functional integral:
\begin{align}
d_{jm}(t) &= d^{\prime}_{jm}(t) - \sum_{m^{\prime},n^{\prime}} G_{m,m^{\prime}}(t) \sigma^3_{m^{\prime},n^{\prime}}  h_{j,n^{\prime}}(t),\\
\bar{d}_{jm}(t) &= \bar{d}^{\prime}_{jm}(t) - \sum_{m^{\prime},n^{\prime}}  \bar{h}_{j,m^{\prime}}(t) \sigma^3_{m^{\prime},n^{\prime}} G_{n^{\prime},m}(t)
\end{align}
where we have denoted $h_{j,m} = -J (c_{j+1,m}c_{j,m} + c_{j,m}c_{j-1,m})$, $\bar{h}_{j,m} = -J ( \bar{c}_{j,m}\bar{c}_{j+1,m} + \bar{c}_{j-1,m}\bar{c}_{j,m})$, and
\begin{equation}
G_{mn}(t) = \frac{ -U \sigma^3_{mn} - i\gamma(t) \left(\sigma^{-}_{mn} + \tfrac{1}{2} \sigma^0_{mn}\right) }{U^2 + (\gamma(t)/2)^2}.
\end{equation}
The resulting integral over $\bar{d}^{\prime}_j,d^{\prime}_{j}$ is gaussian, and yields
a constant prefactor to the generating functional in Eq.~\eqref{eq:zvv}. In addition, there is an exponential factor with the  following effective action in the  exponent:
\begin{equation}
S^{\prime}_{\mathrm{eff}}  = - \sum_{j, mn}\int dt \:  \bar{h}_{j,m}(t) \left( \sigma^3 G(t) \sigma^3\right)_{mn}  h_{j,n}(t).
\end{equation}
Note that in the limit where the coupling to the bath vanishes and $\gamma(t) = 0$, $G_{mn}(t)=-\sigma^{3}_{mn}/U$ and we obtain
\begin{align}
S^{\prime}_{\mathrm{eff}}  &= \frac{1}{U}\sum_{j, mn}\int dt \:  \bar{h}_{j,m}(t) \sigma^3_{mn}  h_{j,n}(t) \\
&= \frac{J^2}{U} \sum_{j,mn}\int dt \: \sigma^3_{mn} \big[ 2 \bar{c}_{j+1,m}\bar{c}_{j,m}c_{j,n}c_{j+1,n} \notag\\ 
&\qquad \qquad - \bar{c}_{j+1,m} \bar{c}_{j,m} c_{j,n} c_{j-1,n} \notag \\ 
&\qquad \qquad -\bar{c}_{j-1,m} \bar{c}_{j,m} c_{j,n} c_{j+1,n}  \big],  
\end{align}
which is the Keldysh action for the effective Hamiltonian obtained using strong coupling perturbation theory in Ref.~\cite{Cazalilla_PhysRevA.67.053606} for the 1D Bose-Hubbard model in the limit where $U\gg J$ and in the subspace with no doublons. 

Next, we switch on the coupling to the bath so that $\gamma(t) = \gamma \theta(t)$ and 
compute the particle-loss rate. It is convenient to work in the Bloch wave basis where $c_k = \sum_{j} e^{-i k x_j} c_{j}/ \sqrt{M}$, $M$ being the number of lattice sites,  $x_j = j$, $k = 2\pi l/M$, $l=-M/2+1, \ldots, M/2$ (assuming periodic boundary conditions and $M$ to be even). Thus, the full effective action $S_{\mathrm{eff}} = S_c+ S^{\prime}_{\mathrm{eff}}$ reads:
 \begin{align}\label{eq:BH2}
& S_{\text{eff}}= \sum_{k,mn}\int dt\,  \sigma^3_{mn} \bar{c}_{k,m} \left( i\partial_{t}-\epsilon_k \right)c_{k,n}\notag\\
&\quad -\frac{1}{2M}\sum_{pkq,mn} \int dt \,  \sigma^3_{mn} \tilde{U}_{pkq}(t) 
\bar{c}_{p,m}\bar{c}_{k,m}c_{k+q,n}c_{p-q,n}\notag \\
 &\quad-\frac{i}{M} \sum_{pkq} \int dt \,  \Gamma_{pkq}(t)  \bar{c}_{p,-}\bar{c}_{k,-}c_{k+q,+}c_{p-q,+}\, , 
\end{align}
where  $\epsilon_k = -J \cos k$ is the single-particle dispersion, $\tilde{U}(p,k,q) = U_{pkq}(t) - i\sigma^3_{mn}\Gamma_{pkq}(t)$ and $\Gamma_{pkq}(t)$ are given by:
\begin{align} 
U_{pkq}(t)&=\frac{-8J^2  U   F_{pkq}}{U^2+(\gamma(t)/2)^2} \\
\Gamma_{pkq}(t)&= \frac{4 J^2 \gamma(t) F_{pkq} }{U^2+(\gamma(t)/2)^2} ,\\
F_{pkq} &= \cos\left(q \right)\cos^2\left(\frac{p+k}{2}\right).
\end{align}
In the limit where $J\ll \max\{U,\gamma(t)\}$, both $U_{pkq}(t)$ and $\Gamma_{pkq}(t)$ are perturbatively small.  Using perturbation theory to leading order in $U_{pkq}(t)$  and $\Gamma_{pkq}(t)$  we obtain  the following loss rate equation for the distribution function of Jordan-Wigner fermions (see Appendix~\ref{app:dev} for further details),
\begin{align}\label{eq:bdloss}
\frac{dn_p(t)}{dt}
&=-\gamma_{\mathrm{eff}} \int \frac{dk}{2\pi}\: C^2_{kp}\: n_{p}(t)n_{k}(t).
\end{align}
where  $C_{kp} =  \sin\left(\frac{p-k}{2}\right)\cos\left(\frac{p+k}{2}\right)$ and the effective loss rate is
\begin{equation}
\gamma_{\mathrm{eff}} = \frac{16J^2\gamma}{U^2+\gamma^2/{4}}.
\end{equation}
Integrating   over  $p$ yields the following rate equation for the lattice 
filling $n_A(t) = N_A(t)/M$:
\begin{align}\label{eq:losseq}
  \frac{d n_A(t)}{dt}&=  -\gamma_{\mathrm{eff}}\int \frac{dpdk}{(2\pi)^2} \: C^2_{kp} \: n_{p}(t)n_{k}(t).
\end{align}
The form factor   $C^2_{kp}$ was not present in the approach used in Ref.~\onlinecite{Garcia-Ripoll_2009}, which neglected inter-site correlations. The expression containing the form factor was later obtained by solving   the Lindblad master equation  using an approximation termed as time-depependent generalized Gibbs
ensemble~\cite{Rossini_PhysRevA.103.L060201,perfetto2022reactionlimited}.  As illustrated below,  
the form factor turns out very important when the filling of the lattice is low and at low initial temperatures. The phenomenological loss rate equation  and the two-body loss coefficient $\gamma_2$ (cf. Eq.~\ref{eq:phenon})~\cite{Syassen_science_2008,Tomita_PhysRevA_2019,Garcia-Ripoll_2009} can be obtained in the high-temperature limit where the Fermi-Dirac distribution of the Jordan-Wigner fermions approaches $n_A(t)$ and is independent of the momentum. Therefore,
\begin{equation}
\frac{dn_A(t)}{dt}=- \gamma_T \: n_A^2(t),\label{eq:phenon2}
\end{equation}  
with two-body loss coefficient $\gamma_2 = \gamma_T=\gamma_{\mathrm{eff}}/4$.
\begin{figure}[t]
\includegraphics[width=1\columnwidth]{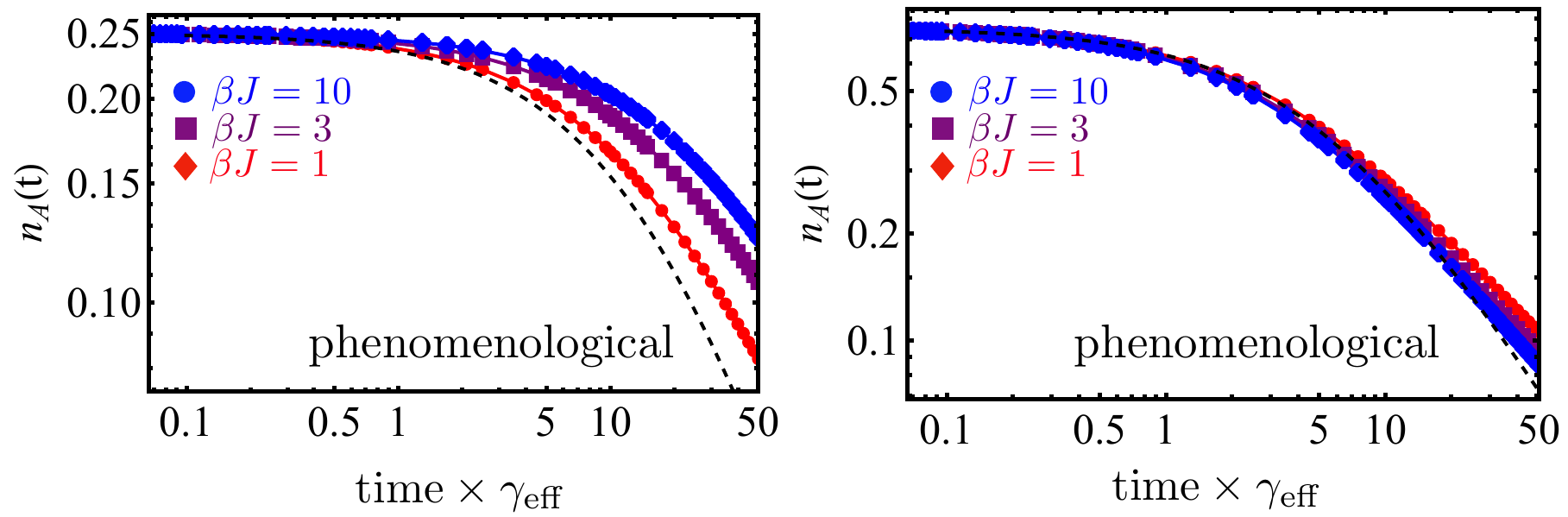}
\center
\caption{ 
(Left) Evolution of the lattice filling starting from an initial filling of $n_A(t = 0)=0.25$ for different initial temperatures on a log-log scale. The initial temperature is determined by the initial state, which is assumed thermal. The phenomenological result shows large deviations from the microscopic theory, especially at low temperatures.
(Right) Evolution of the lattice filling starting from initial filling of $n_A( t= 0)=0.75$ for different initial temperatures on a log-log scale. At larger initial fillings, the deviation from the phenomenological rate equation becomes smaller.}\label{fig:onsite_loss}
\end{figure}
\begin{figure}[t]
\includegraphics[width=1\columnwidth]{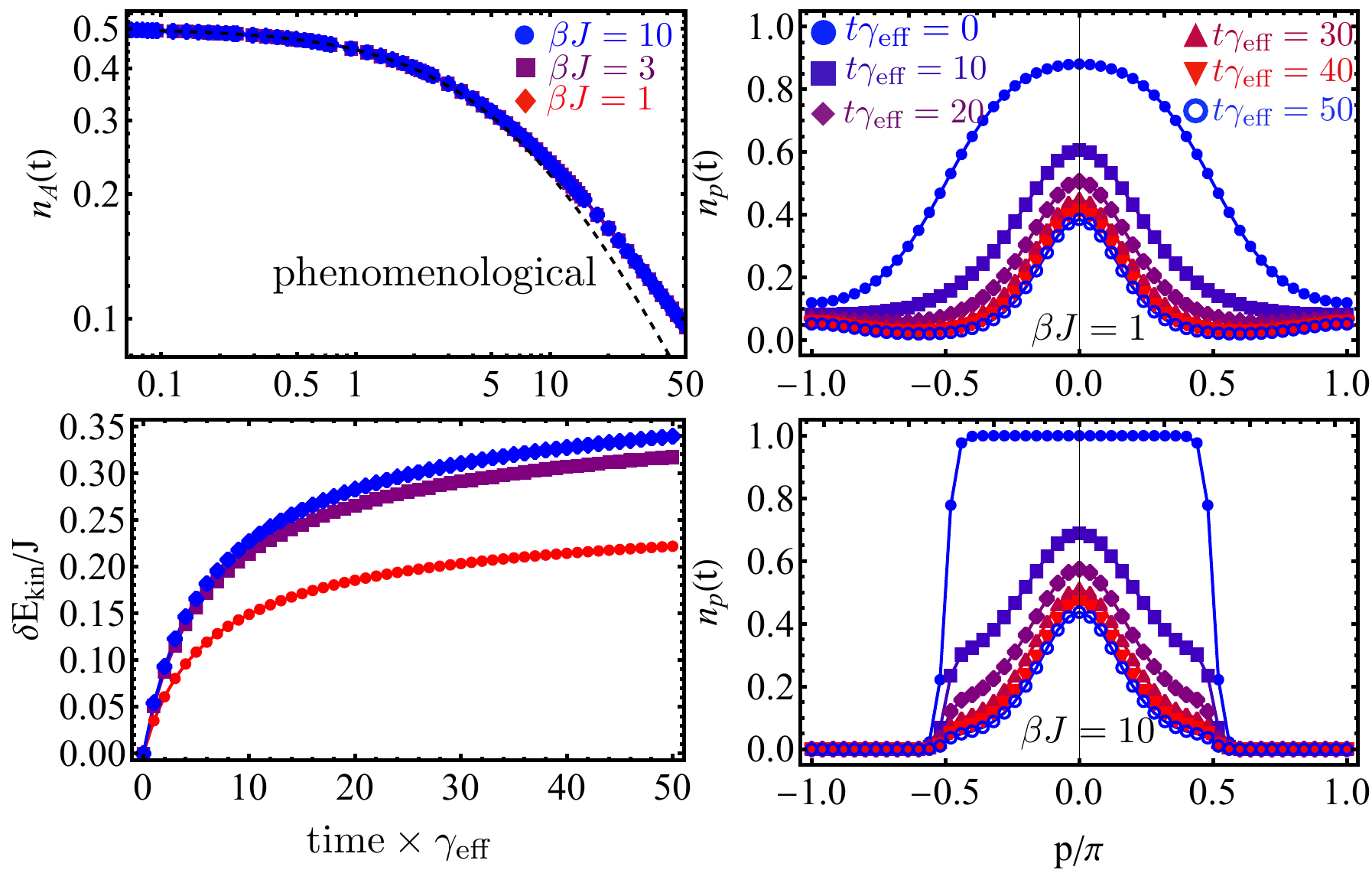}
\center
\caption{
(Top left panel) Decay of the lattice filling for an initially half-filled system and different initial temperatures. The dependence of the initial temperature is very weak (note the log-log scale). (Bottom left panel) change of kinetic energies with different initial temperatures. Systems with a lower initial temperature undergo a 
larger change in the kinetic energy.  (Upper right panel) Time evolution of the distribution of Jordan-Wigner fermions for an inverse temperature  $k_B T = J$ (i.e. $\beta J = 1$). Notice that a large depletion in the distribution near $k = \pm \pi/2$. (Bottom right panel) Evolution of the  distribution of Jordan-Wigner fermions for $k_B T = J/10$ (i.e. $\beta J =10$). }\label{fig:ms}
\end{figure}
Fig.~\ref{fig:onsite_loss}  compares the solution of the phenomenological rate equation with the numerical solution of Eq.~\eqref{eq:bdloss} obtained from the microscopic Keldysh action at different initial temperatures determined by the initial momentum distribution of the Jordan-Wigner fermions. Indeed, since after $t=0$ the system is out of equilibrium, a global temperature can no longer be defined. At a small values of the initial lattice filling (e.g. $n_A(t=0)=0.25$),  the results from the phenomenological rate equation strongly deviate from the predictions of the microscopic theory, Eq.~\eqref{eq:bdloss}.  The 
difference between the two rate equations becomes smaller as temperature increases becasue, as mentioned above, the phenomenological rate equation is recovered from Eq.~\eqref{eq:bdloss} in the high temperature limit where $T \gg J$. However, at a higher initial lattice filling  ($n_A(t=0)=0.75$), the error incurred by using the phenomenological rate 
equation becomes smaller at all the studied temperatures. Thus, the phenomenological rate equation only applies at  high temperatures or high lattice fillings (i.e. $n_A(t=0)\simeq 1$). The latter are indeed the conditions  of  the previous experiments~\cite{Syassen_science_2008,Tomita_PhysRevA_2019}. Our formalism thus allows to access other regimes of  lattice fillings and temperature, which can be explored in future experiments. 

In addition, a closer examination of the numerical solution of Eq.~\eqref{eq:bdloss} shows that 
the situation is indeed more complex than what can be naively  inferred from the  above discussion of the validity of the phenomenological rate equation. In order to 
see from where the complexity emerges, we have  plotted the evolution of the distribution function of Jordan-Wigner fermions  in  Fig.~\ref{fig:ms}  for an initially half-filled lattice and different values of the effective loss rate $\gamma_{\mathrm{eff}}$  (right panels). In the same figure, we also show the evolution of the lattice filling and kinetic energy for different initial temperatures (left panels). Note that for the lattice filling shown on the top left panel, the particle loss dynamics is largely independent of the initial temperature (note the log-log scale). On the other hand, the dynamics of the average kinetic energy (shown in a linear scale) does indeed depend on the initial temperature. As it can be seen, the change in kinetic energy induced by the two-body loss is larger in systems with lower initial temperature. 

 The dependence  on the initial temperature of the kinetic energy, which is the first moment of the momentum distribution, is an indication that the two-body loss drives the system into a non-equilibrium state. To confirm this observation, we focus on the evolution of the full momentum distribution of the Jordan-Wigner fermions, which is shown on the right panels for two values of the initial temperature ($k_B T = J$ and $k_B T = J/10$). Although the initial distribution is assumed to be thermal, it can be seen that  under the two-body loss in both cases it rapidly evolves into a non-thermal distribution. Note that the depletion is most effective (especially
at low temperatures) for $k$ near $\pm \pi/2$. This is because the
form factor $C^2_{kp}$ in Eq.~\eqref{eq:losseq} is maximum for $k = -p = \pm \pi/2$,
which corresponds to losses of doublons with total zero momentum. The latter
are created from  ``Cooper pairs''  $\sim c_{k} c_{-k}$ of Wigner fermions.
In conclusion, even
in cases where evolution of the particle filling stays rather close to the results
obtained from the phenomenological rate equation~\eqref{eq:phenon2}  down to low temperatures, the system is indeed far from equilibrium as revealed by  close examination of other observables like the kinetic energy.

\section{Losses in Optical Feshbach Resonance}\label{sec:ofr}

\begin{figure}[b]
\includegraphics[width=0.99\columnwidth]{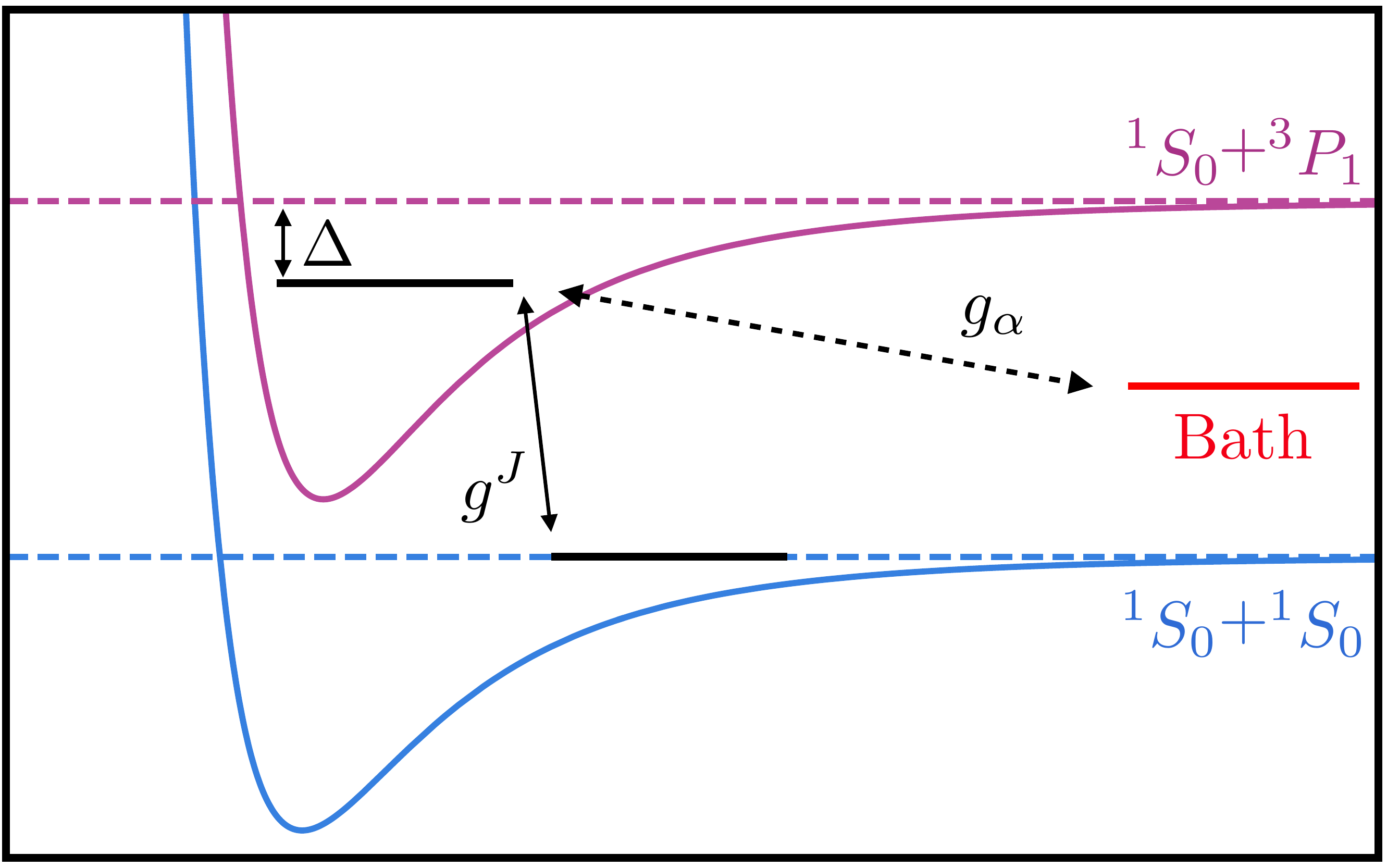}
\center
\caption{Scheme of an optical Feshbach resonance: The system  consists  of $N$ species of spin-$F$ ($N = 2F+1$) alkaline-earth atoms in the ground state interacting via an s-wave potential. Using a laser colliding pairs of atoms in the ground-state are coupled to  pairs consisting of one ground-state  and one excited-state atom in a molecular state. 
This state has a one-body coupling parametrized by $g_{\alpha}$ to a bath to which it can be lost or gained.}
\end{figure}
 In this section we describe a multi-component mixture of alkaline-earth atoms with
emergent SU($N$) symmetry~\cite{Cazalilla_2014,Cazalilla_2009} near an  optical Feshbach resonance (OFR)  by means 
of the following model:
\begin{align}
H_{c}&= H_0 + H_{\mathrm{int}},\\
H_{\mathrm{int}} &= \frac{U}{\Omega}\sum_{\vec{pkq},\sigma\sigma^{\prime}} c^{\dag}_{\vec{p}\sigma}
c^{\dag}_{\vec{k}\sigma^{\prime}} c_{\vec{k-q}\sigma^{\prime}}c_{\vec{p+q}\sigma},\\
H_{ca} &= \frac{f(t)}{\sqrt{\Omega}}\sum_{\vec{pq}} \sum_{\sigma\sigma^{\prime},JM} g^J_{\vec{q}} \langle 
ff\sigma\sigma^{\prime}|JM\rangle\notag\\ &\qquad \times \left[ a^{\dag}_{\vec{p},JM} c_{\vec{p}-\vec{q},\sigma} c_{\vec{q},\sigma^{\prime}}+ c^{\dag}_{\vec{q},\sigma} c^{\dag}_{\vec{p}-\vec{q},\sigma^{\prime}} a_{\vec{p},JM} \right],
\end{align}
\begin{align}
H_a&= \sum_{\vec{q},JM}(\epsilon^a_{\vec{q}} +\Delta) a^\dag_{\vec{q},JM}a_{\vec{q},JM},\\
H_{aB} &=   f(t) \sum_{\vec{q},JM,\alpha}    \left[ g_{\alpha} b^{\dag}_{\vec{q}\alpha}a_{\vec{q},JM} +  g^*_{\alpha} a^\dag_{\vec{q},JM}b_{\vec{q},\alpha} \right],\\
H_\text{B}&=  \sum_{\vec{q},\alpha} \omega_{\vec{q}\alpha}\: b^\dag_{\vec{q}\alpha}b_{\vec{q}\alpha},
\end{align}
Fermions in the ground state  ${^1S_0}$ interact via a weak s-wave potential preserving $SU(N)$ symmetry. The OFR is described by a SU($N$) symmetry breaking scattering channel which involves an intermediate bosonic
molecular state  where one of the  colliding atoms is in an optically coupled excited state, i.e. $^1S_0$ + $^{3}P_0$. The  coupling $g^J(q)=g^J(-q)$ is the matrix element of the laser-induced transition between two-particles in the ground state and the molecular excited state, i.e. $\langle {^1S_0} {^3P_0}|V_\text{las}|{^1S_0} {^1S_0}\rangle$ . In the calculations below, we assume that  $g^J\ll \Delta_m,\gamma$, where  $\Delta_m$ is the detuning from the excited state and $\gamma$ is the one-body loss rate of the excited molecular state. In this limit, the loss of the molecular state is due to spontaneous emission. The contribution of stimulated emissions is relatively small. The two-body loss of the system particles is described as the one-body loss of the intermediate molecular state, which is coupled to a bath via $H_\text{aB}$ whose eigenmodes are the bosonic operators $b^{\dag}_{\vec{q}}, b_{\vec{q}}$. Since the coupling is linear and the bath is described by a quadratic Hamiltonian, the 
second-order cumulant expansion is exact and applies even in the limit where the loss rate $\gamma$ is 
large. Below we write the Keldysh action after integrating out the bath field following the same steps 
than in Sec.~\ref{sec:formalism}. This yields:
\begin{align}
&S_{\text{eff}}=S_c + S_{a,\mathrm{eff}},\\
&S_c= \int dt\, \sigma^3_{mn}  \biggl\{ \left[\sum_{\vec{k}} \bar{c}_{\vec{k}\sigma m}\left( i\partial_{t}-\epsilon_{\vec{k}}  \right) c_{\vec{k}\sigma n}  - H_{\mathrm{int}}\right]\notag\\
       &-  \frac{f(t)}{\sqrt{
    \Omega}}\sum_{\vec{p},\vec{q}} \sum_{\sigma \sigma^{\prime} JM} g^J_{\vec{q}} \langle ff\sigma\sigma^{\prime}|JM\rangle \left[  \bar{a}_{\vec{p} m} c_{\vec{p}-\vec{q}\sigma n}c_{\vec{q}\sigma^{\prime}n} \right. \notag \\
    & \qquad\qquad\qquad\left. +a_{\vec{p} m} \bar{c}_{\vec{p}-\vec{q}\sigma n} \bar{c}_{\vec{q}\sigma^{\prime} n} \right] \biggr\},\notag \\
& S_{a,\mathrm{eff}} =    \sum_{\vec{q},JM} \int dt  dt^{\prime}\, 
\bar{a}_{\vec{q} JM m} G^{-1}_{\vec{q},mn}(t,t^{\prime}) a_{\vec{q} JM n}\,,
\label{eq:Sm'}
\end{align} 
where
\begin{multline} 
G^{-1}_{\vec{q},mn}(t,t^{\prime}) = \delta(t-t^{\prime}) \left[ 
\sigma^{3}_{nm}  \left(i\partial_{t^{\prime}} - \epsilon^a_{\vec{q}} - \Delta \right) \right. \\
\left. + i \sigma^0_{mn}\gamma(t^{\prime})/2 - i\sigma^{-}_{mn}\gamma(t^{\prime}) \right] \label{eq:ginv}
\end{multline}
In Eq.~\eqref{eq:Sm'}, in order to lighten the equations, we have adopted the convention that summation over
repeated indices $m,n =\pm$ is implied.

Next, we integrate out the molecular field, $\bar{a}_{\vec{q},JM}, a_{\vec{q},JM}$ using the 
 cumulant expansion. To this end, we need to obtain the Green's function for the molecular fields, i.e.
 the inverse of the matrix $G^{-1}_0(t,t^{\prime})$. In order to take into account the distribution function
 of the molecules correctly, it is necessary~\cite{kamenev_2011} to take a step back and work with the
 discrete version of the path integral. However, the presence of the dissipative terms make the inversion
 of the matrix cumbersome. One way around this difficulty is to realize that in the parameter regime of interest here, i.e. for large $\gamma(t)$ and/or large detuning $\Delta_m$, it is possible to neglect the time derivative
 part of $G^{-1}_{\vec{q}}(t,t^{\prime})$, which effectively amounts to a Markovian approximation when the
 molecular field is regarded as a bath itself. In order to understand this, let us neglect the time dependence of the coupling to the bath $\gamma$ for the time being. Thus, the matrix $G^{-1}_{\vec{q}}$ becomes a function of $t-t^{\prime}$ and it can be Fourier transformed, which yields the following
 matrix in Keldysh space:
\begin{align}
G^{-1}_{\vec{q}}(\omega) =  \sigma^3 \left(\omega - \epsilon^a_{\vec{q}} - \Delta \right)
+ i \gamma \left( \sigma^0/2 - \sigma^{-} \right).
\end{align}
If $|\omega|\ll \min\{\Delta_m,\gamma\}$ we can neglect the frequency dependence of 
$G^{-1}_{\vec{q}}(\omega)$. Furthermore, assuming a vanishing number of molecules in the initial state,
the correlation functions that determine the second order term in the expansion are given
by the inverse of above matrix with $\omega = 0$. In Eq.~\eqref{eq:ginv} this amounts to
neglecting the term involving the time derivative, $i\partial_t$, which yields Markovian
correlations for the molecular field of the form:
\begin{multline}
G_{\vec{q},m,n}(t,t^{\prime}) \simeq  -\frac{\delta(t-t^{\prime})}{(\epsilon_{\vec{q}}+\Delta)^2+ \gamma^2(t^{\prime})/4}\\
\times \left[ \sigma^{3}_{nm}  \left(\epsilon^a_{\vec{q}} + \Delta \right)  + \tfrac{i}{2}\left( \sigma^0_{mn} + \sigma^{-}_{mn} \right) \gamma(t^{\prime}) \right]
\end{multline}
Using the above expression and Eq.~\eqref{eq:S2eff}, we obtain the  following effective Keldysh action:
\begin{align}\label{eq:fesh2}
 S_{\text{eff}}&= \sum_{\vec{k},\sigma} \int dt \biggl\{ \sigma^3_{mn} \bar{c}_{\vec{k}\sigma m}\left( i\partial_{t}  - \epsilon_{\vec{k}} \right) c_{\vec{k}\sigma n} \notag\\
&- \frac{1}{2\Omega}\sum_{\vec{p}\vec{k}\vec{q}} \sum_{\sigma \sigma^{\prime}\lambda\lambda^{\prime}, JM} 
\biggl[ \sigma^3_{mn} U_{\text{eff}}(\vec{p},\vec{k},\vec{q}) - i \sigma^0_{mn} \gamma^{\prime}(\vec{p},\vec{k},\vec{q}) \biggr] \notag\\
& \qquad \times \bar{c}_{\vec{p}\sigma m}\bar{c}_{\vec{k}\sigma^{\prime} m}c_{\vec{k+q} \lambda n}
c_{\vec{p-q},\lambda^{\prime},n} \biggr\},\notag \\
-i &   \int dt  \sum_{\vec{p},\vec{k},\vec{q}} \sum_{\sigma \sigma^{\prime}\lambda\lambda^{\prime}, JM}  
\frac{\gamma^{\prime}(\vec{p},\vec{k},\vec{q},t)}{\Omega}, \notag \\&
\qquad\times  \bar{c}_{\vec{p}\sigma -}\bar{c}_{\vec{k}\sigma^{\prime} -}
c_{\vec{k+q}\lambda+} c_{\vec{p-q}\lambda^{\prime}+},
\end{align}
where $U_{\text{eff}}(\vec{p},\vec{k},\vec{q},t)=U_0+\delta U(\vec{p},\vec{k},\vec{q},t)$ 
is the renormalized interaction: 
\begin{multline}
       \delta U(\vec{p},\vec{k},\vec{q},t) =-2  
    g^J_{\vec{p}-\vec{k}} g^J_{\vec{p}-\vec{k}-2\vec{q}}\langle ff\sigma\sigma'|JM\rangle\\
  \qquad\qquad  \times  \langle ff\lambda\lambda'|JM \rangle
    \frac{\Delta_{\vec{p}+\vec{k}} f^2(t) }{\Delta^2_{\vec{p}+\vec{k}}+ \tfrac{\gamma^2(t)}{4}},
\end{multline}
%
%
and  
\begin{multline}
\gamma^{\prime}(\vec{p},\vec{k},\vec{q},t)=g^{J}_{\vec{p}-\vec{k}}g^J_{\vec{p}-\vec{k}-2\vec{q}} 
\langle ff\sigma\sigma^{\prime}|JM\rangle \\ 
\times   \langle ff\lambda\lambda^{\prime}  |JM\rangle \frac{\gamma(t) f^2(t)}{\Delta^2_{\vec{p}+\vec{k}}+\frac{\gamma^2(t)}{4}}  
\end{multline}
is the effective two-body loss rate in the limit of strong spontaneous loss on the intermediate molecular states~\cite{Thies_PhysRevLett.93.123001,Ciurylo_PhysRevA_2004,enomoto_PhysRevLett.101.203201,Napolitano_PhysRevLett.73.1352,Yamazaki_PhysRevA.87.010704,Kim_PhysRevA.94.042703,Bohn_PhysRevA.56.1486}. In the above expressions $\Delta_{\vec{p}}= \epsilon^a_{\vec{p}}+\Delta$ is the energy of the excited molecular state with total momentum $\vec{p}$. Notice that $\delta U$ is an $SU(N)$-symmetry breaking interaction and both  quantities are perturbatively small in the limit where $g^J\ll \min\{\Delta_m,\gamma\}$ of interest here. Hence, perturbation theory to leading order in $\gamma^{\prime}$ yields the following rate equation for an quantum degenerate gas:
\begin{align}
\frac{dn_c(t)}{dt}= -\frac{2}{\Omega^2}\sum_{\vec{p},\vec{k}}\gamma^{\prime}(\vec{p},\vec{k},\vec{0})n(\vec{p},t)n(\vec{k},t).
\end{align}
where $n_{\vec{p}}(t)$ is the instantaneous momentum distribution and $n_c=\sum_\vec{p} n_\vec{\vec{p}}(t)/\Omega$ is the fermions density. From this result, the phenomenological loss coefficient for a thermal gas can be obtained by replacing the loss-coefficient with its thermal average~\cite{RevModPhys.82.1225, Ciurylo_PhysRevA_2004,Blatt_PhysRevLett_2011},
\begin{align}
\gamma_T&=\langle \gamma^{\prime}(\vec{p},\vec{k},0) \rangle_T,\\
&=\int \gamma^{\prime}(\vec{p},\vec{k},0) f_M(\vec{p},T) f_M(\vec{k},T) d^3\vec{p} d^3\vec{k}.
\end{align}
Here $f_M(\vec{p},T)=(2\pi m k_B T )^{-3/2}\exp(-p^2/2m k_B T)$ denotes the  Maxwell distribution at temperature $T$ for particles with mass $m$ normalized to unity. Using the average coefficient, the rate equation can be approximated by
\begin{align}
\frac{d n_c(t)}{dt}&= -\gamma_T\: \frac{2}{\Omega^2}\sum_{\vec{p},\vec{k}}n(\vec{p},t)n(\vec{k},t),\\
&=-2\gamma_T \: n_c^2(t),
\end{align}
which is the phenomenological two-body loss rate equation, Eq.~\eqref{eq:phenon} with $\gamma_1 = 0$ and $\gamma_2 = 2\gamma_T$ for a thermal gas~\cite{RevModPhys.82.1225,Ciurylo_PhysRevA_2004,enomoto_PhysRevLett.101.203201,Blatt_PhysRevLett_2011,Yamazaki_PhysRevA.87.010704}.

\section{conclusion}

 In this work, we have discussed the derivation of the Keldysh path integral
for open quantum systems using the 2nd order cumulant expansion.
Although  we have focused on Markovian baths, the method
is not limited to the latter and can be extended to describe effects
beyond Markovianity. It also does not require the bath to be non-interacting
or the bath coupling to be of a particular form. Formally, it requires that the
coupling  to the bath is weak enough to be accurately treated using second order
perturbation theory. However, as we have shown above, when describing 
two-particle losses, the system-bath coupling can be sometimes conveniently
reformulated and a strong loss regime can be also described.

Turning to models relevant for ultra-cold atomic gases, we have studied  models of one and  two-body losses. Thus, we have shown how
two-body losses caused by photo-association of doublons in the one dimensional Bose-Hubbard model can be described within the path-integral formulation allowing us to obtain a microscopic loss
rate equation. The latter has been compared with a previously derived phenomenological loss equation. The microspically derived rate equation shows that the phenomenological equation is mostly accurate at  high temperatures  and/or lattice fillings close to unity.  However, we have shown (see Sect.~\ref{sec:lossybhm}) that even in  cases where the phenomenological rate equation appears to be sufficiently accurate, the implicit assumption 
that the system remains in a thermal equilibrium characterized by a temperature $T$ can be incorrect. 
This has important implications for the calculation of other physical quantities such as the kinetic energy, 
for which our theory, which properly handles such deviations from equilibrium, is necessary. 

Finally, in Sec.~\ref{sec:ofr} we have applied the formalism to a model describing an
optical Feshbach resonance in multi-component mixture of ultracold alkaline-earth fermions. We have
also shown that the phenomenological rate equation is expected to apply to the high temperature
regime.  Although we have not fully explored the low temperature regime yet, 
using the lessons learned with the much simpler one-dimensional Hubbard model, in presence of an OFR 
we expect that if quantum coherence becomes important, deviations from the phenomenological approach will appear. 

Finally, let us mention that, in this work, when dealing with the interactions between the Jordan-Wigner fermions in the lossy one-dimensional Bose-Hubbard model we have used perturbation theory. This is justified because the 
 latter are weak~\cite{Cazalilla_PhysRevA.67.053606} and the studied temperatures are relatively high.
 However, in future work it will interesting to revisit this problem in order to account for the effect
 of the interactions in the system, which in one dimension can be done non-perturbatively using 
 bosonization~\cite{Giamarchi_1dbook,Cazalilla_2004}. Other possible extensions of this
 work are, as pointed out above, studying effects beyond Markovianity and the effect 
 of strong interactions/correlations in the bath.

\section*{Acknowlegements}

 MAC acknowledges support from Ikerbasque, Basque Foundation for Science. In addition, CHH and MAC  have been supported by  MCIN Grant No. PID2020-120614GB-I00 (ENACT). CHH also acknowledges a PhD fellowship from DIPC 
This work was also supported in part by the Swiss National Science foundation under Division II (grant 200020-188687). 
TG would like to thank S. Uchino and J.P. Brantut for interesting discussions. 
 
\appendix

\section{Brief recap of the path integral approach to the Lindblad master equation}\label{app:recap}

For an open quantum system described by a Hamiltonian $H = H(a^{\dag},a)$ in contact with 
a Markovian bath, the time-evolution of the  reduced density matrix of the system $A$,
$\rho_A$, reads:
\begin{equation}
\partial_t \rho_A = -i \left[H,\rho_A\right] +\sum_{s} \gamma_{s} \left(L_s \rho_s L^\dag_{s} - \frac{1}{2}\{L^\dag_{s}L_s ,\rho_A  \} \right),\label{eq:master}
\end{equation}
where $[O_1,O_2] = O_1 O_2-O_2 O_1$ and $\{O_1,O_2\}=O_1 O_2+ O_2 O_1$ and the couplings $\gamma_{\alpha}$ along with the Lindblad operators $L_{\alpha}= L_{\alpha}(a^{\dag},a)$ and $L^{\dag}_{\alpha}= L^{\dag}_{\alpha}(a^{\dag},a)$ describe the coupling to the  bath. Specializing to the case where the Lindbladian  describes particle losses (but not gains), and using the resulting time-evolution operator derived from the above master equation, a path integral can be derived following the procedure described in Ref.~\cite{Sieberer2016} (see equations 27 and 28 in  Ref.~\cite{Sieberer2016}). In the notations of this manuscript where $\psi_{\pm} \to a_{\pm}$ and $-i\mathcal{L}\to \mathcal{L}$ describes only the dissipative part of the Keldysh action, et cetera, the path integral derived in Ref.~\cite{Sieberer2016} reads:
\begin{align}
Z&=\int D[a_+,\bar{a}_+,a_-,\bar{a}_-]\, e^{i S},\\
S&=\int dt \left[ \sigma_{mn}^3\bar{a}_m i\partial_t a_n   - \sigma_{mn}^3 H( \bar{a}_m,a_n) \right. \notag\\
&\qquad\qquad\left. + \mathcal{L}(\bar{a}_+,\bar{a}_-,a_{+},a_{-})\right],\\
\mathcal{L} &= -i \sum_s \gamma_{s}\left[\bar{L}_{s,-} L_{s,+} -\tfrac{1}{2}\left(\bar{L}_{s,+}L_{s,+}+\bar{L}_{s,-}L_{s,-}\right)\right].
\end{align}
Note that the ``quantum jump'' term $\bar{L}_{s,-} L_{s,+}$ appears in the reversed order compared
to the Lindblad master equation (i.e. $L_s\rho L^{\dag}_s$ in Eq.~\ref{eq:master}). This order is important in the fermion case
when the Lindblad operator $L_s$ contains an odd number of fermion operators~\cite{TONIELLI_2016}.
In order to make contact with our approach, we shall consider the case of one-particle losses studied in Secs.~\ref{sec:singlemode} and \ref{sec:onebody} for which  the Lindbladian operators are $L_{s} = a$ and $\gamma_{s}  = \gamma(t)$ and $H(a^{\dag},a) = \epsilon_0\: a^{\dag}a$, for the single-mode  case (i.e. dropping $s$, the generalization to the multiple mode corresponds to letting $s = \vec{k}$ and $\gamma_{s=\vec{k}} = \gamma(t)$ as in Sec.~\ref{sec:onebody}). Thus, $\mathcal{L}$ given above becomes our Eq.~\eqref{eq:single mode}.

\section{2nd order cumulant expansion}\label{app:cumu}

In this section, we provide a short derivation of  the 2nd order cumulant expansion in the context of path integral as used in Sec.~\ref{sec:singlemode}. 
Consider the derivation of the Feynman-Vernon functional which is obtained by formally
integrating out the bath $B$ degrees of freedom, i.e.
\begin{align}
\mathcal{F}[\bar{a},a] &= \langle e^{i S_{AB}} \rangle_B , \notag\\
&= \langle     1+i S_{AB} - \frac{1}{2}  S_{AB}^2   +\cdots \rangle_B \notag, \\
&=  \text{exp} \left [ \log \left  (  \langle1+i S_{AB} - \frac{1}{2}  S_{AB}^2 +\cdots \rangle_B  \right  ) \right ]    \notag, \\
&\simeq  \exp\left[
 i \langle S_{AB}\rangle_B 
 - \frac{1}{2} \left( \langle S^2_{AB} \rangle_B - \langle S_{AB} \rangle_B^2 \right) 
\right]\notag, \\ 
&= e^{i \mathcal{L}[\bar{a},a]}.
\end{align}
where in the third line, we use the expansion $\log(1+x) = \sum_{n=1}^{\infty} (-1)^{n+1}\frac{x^n}{n}$ and kept the terms up to second order in $S_{AB}$.
\section{Bath correlations in dilute limit and Markovian approximation}\label{app:corr}

 The coupling functions $g^{mn}(t_1-t_2)$ are expressed in terms of bath correlators. In the limit of very dilute bath excitations i.e. $n_B(\omega_{\alpha})\approx 0$, 
 the effects of interactions in the bath can be neglected and  the correlation functions are well
 approximated by those of an non-interacting system. Before embarking in the calculation, it is important
 to note that, e.g. the time-ordered correlation takes the form~\cite{kamenev_2011}:
\begin{multline}   
\langle  b_{\alpha+}(t_1) \bar{b}_{\alpha+}(t_2) \rangle_B = \bigg\{ \tilde{\theta}(t_1-t_2) \left[ 1+ z n_B(\omega_{\alpha}) \right]  \\ 
 + \theta(t_2-t_1) z n_B (\omega_{\alpha}) \bigg\} e^{-i \omega_{\alpha}(t_1-t_2)}.
\end{multline}
In the above expression, we have made explicitly the distinction between the two kinds of step functions resulting from the discrete version of the path integral with $\tilde{\theta}(0) = 1$ and $\theta(0) = 0$~\cite{kamenev_2011}. Note that this prescription differs from the one used in the operator approach to Keldysh perturbation theory but in this context this  regularization is dictated by the discrete form path integral, which is  ultimately the mathematically correct form of the latter. As we will show below, the distinction introduced by this regularization is important when taking the Markovian limit where some of the above time arguments become coincident. Furthermore, in the limit where the bath excitations are dilute, the  correlations simplify to:
\begin{align}
 \langle  b_{\alpha+}(t_1) \bar{b}_{\alpha+}(t_2) \rangle_B &= \tilde{\theta}(t_1-t_2) e^{-i \omega_{\alpha}(t_1-t_2)},\\
 \langle  \bar{b}_{\alpha+}(t_2) b_{\alpha+}(t_1) \rangle_B &= z  \theta(t_1-t_2) e^{-i \omega_{\alpha}(t_1-t_2)},\\
 \langle  b_{\alpha-}(t_1) \bar{b}_{\alpha-}(t_2) \rangle_B &= \tilde{\theta}(t_2-t_1) e^{-i \omega_{\alpha}(t_1-t_2)},\\
\langle  \bar{b}_{\alpha-}(t_2) b_{\alpha-}(t_1) \rangle_B &=   z \theta(t_2-t_1) e^{-i \omega_{\alpha}(t_1-t_2)},\\ 
\langle b_{\alpha +}(t_1) \bar{b}_{\alpha-}(t_2) \rangle_B  &= 0,\\
\langle \bar{b}_{\alpha-}(t_1) b_{\alpha +}(t_2) \rangle_{B}  &=0,\\
 \langle b_{\alpha -}(t_1) \bar{b}_{\alpha+}(t_2) \rangle_B  & =  e^{-i\omega_{\alpha}(t_1-t_2)},\\
\langle \bar{b}_{\alpha+}(t_2) b_{\alpha-}(t_1) \rangle_{B} &=   z e^{-i\omega_{\alpha}(t_1-t_2)}.
\end{align}
The effective coupling  are  fully determined by Fourier transform of the following spectral density of couplings to the bath:
 \begin{equation}
\mathcal{J}_B(\omega) =  \sum_{\alpha} |g_{\alpha}|^2 \delta(\omega-\omega_{\alpha}).\label{eq:spectral}
\end{equation}
In terms of the spectral density of couplings to the bath, the functions $g^{\alpha\beta}(t_1,t_2)$ can be written as follows: 
\begin{align}
g^{++}(t_1-t_2) &\simeq   \left\{ \tilde{\theta}(t_1-t_2)  \right.
\notag\\
&+\left.  \theta(t_1-t_2)  \right\} \int \frac{d\omega}{2\pi}  \mathcal{J}_B(\omega) e^{-i\omega(t_1-t_2)},\\
g^{--}(t_1-t_2) &\simeq \left\{ \tilde{\theta}(t_2-t_1)   \right.\notag \\
& + \left.  \theta(t_1-t_2)  \right\} \int \frac{d\omega}{2\pi}  \mathcal{J}_B(\omega) e^{-i\omega(t_1-t_2)},
\end{align}
\begin{align}
g^{-+}(t_1-t_2) &\simeq  -2 \, \int \frac{d\omega}{2\pi}  \mathcal{J}_B(\omega)  e^{-i\omega(t_1-t_2)},\\
g^{+-}(t_1-t_2) &\simeq 0.
\end{align}
By further assuming  that, within a band of width $D$ around $\omega =0$, $\mathcal{J}_B(\omega)$ is well approximated by a constant, i.e.  
\begin{equation}
\mathcal{J}_B(\omega) =   \nu_0 |\langle g\rangle|^2 = \mathrm{const.\,\,},
\end{equation}
for $|\omega| < D/2$,  where $\langle g\rangle$ is the average coupling strength and $\nu_0\sim D^{-1}$ is the density of states. We note, this constant dependence in the spectral density corresponds to a Markovian bath. Within our formalism, non-Markovianity can be introduced using other types of spectral densities, e.g. power-law forms~\cite{Weissbook2008}, which yield  bath correlations different from the Markovian one. In the Markovian case, using the above expression, we have 
\begin{equation}
\int \frac{d\omega}{2\pi}  \mathcal{J}_B(\omega)  e^{- i\omega t} 
\simeq  \nu_0 |\langle g\rangle|^2   \frac{\sin( D t/2 )}{\pi t}.
\end{equation}
Note that this function is strongly peaked for $t  = t_1-t_2= 0$ and decreases/oscillates rapidly for $|t_1-t_2| \gtrsim D^{-1}$ (the oscillation is an artifact of the hard cutoff). Thus, if the dynamics of the system is characterized by frequencies much smaller than $D$, we can effectively replace the above function by a Dirac delta-function $\delta(t_1-t_2)$, which yields,
\begin{align}
v^{++}(t_1,t_2)  &\simeq   \nu_0  |\langle g\rangle f(t_1) |^2   \delta(t_1-t_2) , \\
v^{--}(t_1,t_2)  &\simeq  \nu_0  |\langle g\rangle f(t_1) |^2   \delta(t_1-t_2),\\
v^{-+}(t_1,t_2)&\simeq -2 \nu_0 |\langle g\rangle f(t_1) |^2   \delta(t_1-t_2),\\
v^{+-}(t_1,t_2) &\simeq 0.
\end{align}
In order to obtain the above expressions, we have used $\tilde{\theta}(t) \delta(t) = \tilde{\theta}(0) \delta(t) = \delta(t)$ 
and $\theta(t) \delta(t) = \theta(0)\delta(t) = 0$, as required by the discrete version of the path integral~\cite{kamenev_2011}.
The $\tilde{\theta}(t)$ and $\theta(t)$ are  two regularizations of the step function.

\section{Derivation of loss equation for lossy 1D Bose Hubbard model}\label{app:dev}

The momentum distribution $n_r(t)$ for particle with momentum $r$ at  time $t$ using perturbation expansion to leading order is
\begin{align}\label{eq:m}
&n_{ r }(t) - n_r^0 \simeq i \langle \bar{c}_{r,-}(t) c_{r,+}(t)  \mathcal{L}\rangle_{c},\notag\\
&\notag\\
&=i \biggl\{\frac{-1}{2M}\sum_{pkq,mn} \int_{-\infty}^t dt_1\,    \sigma^3_{mn} U_{pkq}(t_1) O_{mn}(t,t_1;r,p,k,q)\notag\\
&+\frac{i}{2M}\sum_{pkq,mn} \int_{-\infty}^t dt_1\, \sigma^0_{mn}\Gamma_{pkq}(t_1) O_{mn}(t,t_1;r,p,k,q)\notag\\
&-\frac{i}{M} \sum_{pkq} \int_{-\infty}^t dt_1 \,  \Gamma_{pkq}(t_1) O_{-+}(t,t_1;r,p,k,q)
\biggr\},
\end{align}
where  $n_r^0 = \langle c^\dag_{r,-}(t) c_{r,+}(t) \rangle_{c} $ is the momentum distribution of the Jordan-Wigner Fermions in the initial state described by the (non-interacting) Jordan-Wigner fermion action $S_c$ (cf. Eq.~\ref{eq:sc} or first term on the right hand side of Eq.~\ref{eq:BH2}) with initial inverse temperature $\beta$. In the above expression $\langle \ldots \rangle_c$ stands for
Keldysh time-ordered average with weight $e^{iS_c}$. In addition, we have also introduced the following notation for the six fermion-operator expectation values:
\begin{widetext}
\begin{align}
O_{mn}(t,t_1;r,p,k,q) 
&=\langle \bar{c}_{r,-}(t) c_{r,+}(t)\, \bar{c}_{p,m}(t_1) \bar{c}_{k,m}(t_1)c_{k+q,n}(t_1)c_{p-q,n}(t_1)  \rangle_{c}. 
 \end{align}
\end{widetext}
Expanding  the sum including $\sigma^3_{mn}$ and $\sigma^0_{mn}$, we have
\begin{align}\label{eq:O30}
\sum_{mn} &\sigma^3_{mn}O_{mn}(t,t_1;r,p,k,q)\notag\\
&=O_{++}(t,t_1;r,p,k,q)-O_{--}(t,t_1;r,p,k,q).\\
\sum_{mn} &\sigma^0_{mn}O_{mn}(t,t_1;r,p,k,q)\notag\\
&=O_{++}(t,t_1;r,p,k,q)+O_{--}(t,t_1;r,p,k,q).
\end{align}
Next, applying Wick theorem yields:
\begin{align}
& O_{mn}(t,t_1;r,p,k,q)\notag \\
&= \biggl [ \langle \bar{c}_{k,m}(t_1)c_{k+q,n}(t_1)  \rangle_{c}\langle c_{r,+}(t)\, \bar{c}_{p,m}(t_1) \rangle_{c}
\notag\\& \qquad\qquad\qquad\qquad\qquad\qquad\times\langle \bar{c}_{r,-}(t)c_{p-q,n}(t_1)\rangle_{c}\notag \\ &+
\langle  \bar{c}_{p,m}(t_1)c_{p-q,n}(t_1)  \rangle_{c}\langle c_{r,+}(t)\, \bar{c}_{k,m}(t_1) \rangle_{c}\notag\\& \qquad\qquad\qquad\qquad\qquad\qquad\times\langle \bar{c}_{r,-}(t)c_{k+q,n}(t_1)  \rangle_{c} 
\biggr]\notag \\
&-  \biggl[ \langle \bar{c}_{k,m}(t_1)c_{p-q,n}(t_1)  \rangle_{c}\langle c_{r,+}(t)\, \bar{c}_{p,m}(t_1) \rangle_{c}\notag\\& \qquad\qquad\qquad\qquad\qquad\qquad\times\langle \bar{c}_{r,-}(t)c_{k+q,n}(t_1)  \rangle_{c}\notag \\
&+\langle \bar{c}_{p,m}(t_1)c_{k+q,n}(t_1)  \rangle_{c}\langle c_{r,+}(t)\, \bar{c}_{k,m}(t_1) \rangle_{c}\notag\\& \qquad\qquad\qquad\qquad\qquad\qquad\times\langle \bar{c}_{r,-}(t)c_{p-q,n}(t_1)  \rangle_{c}
\biggr],
\end{align}
which yields
\begin{align}\label{eq:OPPMM}
&O_{++}(t,t_1;r,p,k,q)=O_{--}(t,t_1;r,p,k,q)\notag\\
&=\biggl[   \delta_{q,0} \delta_{r,p} n_k^0   + \delta_{q,0} \delta_{r,k} n_p^0   \notag  \\
&-\delta_{q,p-k} \delta_{r,p} n_k^0   - \delta_{q,p-k} \delta_{r,k} n_p^0   \biggr ] \notag \\
&\times \biggl[\tilde{\theta}(t-t_1) \big(1-n_r^0 \big)n_r^0 - \theta(t_1-t) (n_r^0)^2\biggr].
\end{align}
 and
 \begin{align}\label{eq:OMP}
 O_{-+}(t,t_1;r,p,k,q) &= -\biggl[   \delta_{q,0} \delta_{r,p} n_k^0 + \delta_{q,0} \delta_{r,k} n_p^0 \notag\\
&- \delta_{q,p-k} \delta_{r,p} n_k^0  -\delta_{q,p-k} \delta_{r,k} n_p^0\biggr ] (n_r^0)^2.
  \end{align}
  Note that the exponential phase dependence on  $t$ and $t_1$ is cancelled in the above first order expectation values. Finally, combining Eq.~\eqref{eq:m}, \eqref{eq:O30}, \eqref{eq:OPPMM} with Eq.~\eqref{eq:OMP} and re-arranging the momentum indices yields:
 \begin{align} 
n_r(t)-n_r^0 &= -\frac{2}{M}\sum_{k}\left[\Gamma_{rk,q=0}-\Gamma_{rk,q=r-k}\right] n^0_{r}n^0_{k},\notag\\
&=\frac{-1   }{M}\sum_k\:   \int_{-\infty}^t  \frac{16 J^2 \gamma(t_1)}{U^2+\gamma^2(t_1)/4}   \: n_{r}^0n_{k}^0 \: dt_1\notag\\
&\quad\quad\times \sin^2\left(\frac{r-k}{2}\right) \cos^2\left(\frac{r+k}{2}\right).                
\end{align}
Setting $\gamma(t)=\theta(t)\gamma$ and taking $t\to 0$,  we arrive at the loss rate equation given in 
Eq.~\eqref{eq:bdloss}.

\bibliography{lindblad,paper2,paper4,paper-local}

\begin{thebibliography}{67}%
\makeatletter
\providecommand \@ifxundefined [1]{%
 \@ifx{#1\undefined}
}%
\providecommand \@ifnum [1]{%
 \ifnum #1\expandafter \@firstoftwo
 \else \expandafter \@secondoftwo
 \fi
}%
\providecommand \@ifx [1]{%
 \ifx #1\expandafter \@firstoftwo
 \else \expandafter \@secondoftwo
 \fi
}%
\providecommand \natexlab [1]{#1}%
\providecommand \enquote  [1]{``#1''}%
\providecommand \bibnamefont  [1]{#1}%
\providecommand \bibfnamefont [1]{#1}%
\providecommand \citenamefont [1]{#1}%
\providecommand \href@noop [0]{\@secondoftwo}%
\providecommand \href [0]{\begingroup \@sanitize@url \@href}%
\providecommand \@href[1]{\@@startlink{#1}\@@href}%
\providecommand \@@href[1]{\endgroup#1\@@endlink}%
\providecommand \@sanitize@url [0]{\catcode `\\12\catcode `\$12\catcode
  `\&12\catcode `\#12\catcode `\^12\catcode `\_12\catcode `\%12\relax}%
\providecommand \@@startlink[1]{}%
\providecommand \@@endlink[0]{}%
\providecommand \url  [0]{\begingroup\@sanitize@url \@url }%
\providecommand \@url [1]{\endgroup\@href {#1}{\urlprefix }}%
\providecommand \urlprefix  [0]{URL }%
\providecommand \Eprint [0]{\href }%
\providecommand \doibase [0]{https://doi.org/}%
\providecommand \selectlanguage [0]{\@gobble}%
\providecommand \bibinfo  [0]{\@secondoftwo}%
\providecommand \bibfield  [0]{\@secondoftwo}%
\providecommand \translation [1]{[#1]}%
\providecommand \BibitemOpen [0]{}%
\providecommand \bibitemStop [0]{}%
\providecommand \bibitemNoStop [0]{.\EOS\space}%
\providecommand \EOS [0]{\spacefactor3000\relax}%
\providecommand \BibitemShut  [1]{\csname bibitem#1\endcsname}%
\let\auto@bib@innerbib\@empty
\bibitem [{\citenamefont {Breuer}\ \emph {et~al.}(2002)\citenamefont {Breuer},
  \citenamefont {Petruccione} \emph {et~al.}}]{breuer2002}%
  \BibitemOpen
  \bibfield  {author} {\bibinfo {author} {\bibfnamefont {H.-P.}\ \bibnamefont
  {Breuer}}, \bibinfo {author} {\bibfnamefont {F.}~\bibnamefont {Petruccione}},
  \emph {et~al.},\ }\href
  {https://doi.org/10.1093/acprof:oso/9780199213900.001.0001} {\emph {\bibinfo
  {title} {The theory of open quantum systems}}}\ (\bibinfo  {publisher}
  {Oxford University Press},\ \bibinfo {year} {2002})\BibitemShut {NoStop}%
\bibitem [{\citenamefont {Daley}(2014)}]{daley_quantum_2014}%
  \BibitemOpen
  \bibfield  {author} {\bibinfo {author} {\bibfnamefont {A.~J.}\ \bibnamefont
  {Daley}},\ }\bibfield  {title} {\bibinfo {title} {Quantum trajectories and
  open many-body quantum systems},\ }\href
  {https://doi.org/10.1080/00018732.2014.933502} {\bibfield  {journal}
  {\bibinfo  {journal} {Advances in Physics}\ }\textbf {\bibinfo {volume}
  {63}},\ \bibinfo {pages} {77} (\bibinfo {year} {2014})}\BibitemShut {NoStop}%
\bibitem [{\citenamefont {Ashida}\ \emph
  {et~al.}(2020{\natexlab{a}})\citenamefont {Ashida}, \citenamefont {Gong},\
  and\ \citenamefont {Ueda}}]{ashida_non-hermitian_2020}%
  \BibitemOpen
  \bibfield  {author} {\bibinfo {author} {\bibfnamefont {Y.}~\bibnamefont
  {Ashida}}, \bibinfo {author} {\bibfnamefont {Z.}~\bibnamefont {Gong}},\ and\
  \bibinfo {author} {\bibfnamefont {M.}~\bibnamefont {Ueda}},\ }\bibfield
  {title} {\bibinfo {title} {Non-{Hermitian} physics},\ }\href
  {https://doi.org/10.1080/00018732.2021.1876991} {\bibfield  {journal}
  {\bibinfo  {journal} {Advances in Physics}\ }\textbf {\bibinfo {volume}
  {69}},\ \bibinfo {pages} {249} (\bibinfo {year}
  {2020}{\natexlab{a}})}\BibitemShut {NoStop}%
\bibitem [{\citenamefont {M\"uller}\ \emph {et~al.}(2012)\citenamefont
  {M\"uller}, \citenamefont {Diehl}, \citenamefont {Pupillo},\ and\
  \citenamefont {Zoller}}]{MuellerZoller2012}%
  \BibitemOpen
  \bibfield  {author} {\bibinfo {author} {\bibfnamefont {M.}~\bibnamefont
  {M\"uller}}, \bibinfo {author} {\bibfnamefont {S.}~\bibnamefont {Diehl}},
  \bibinfo {author} {\bibfnamefont {G.}~\bibnamefont {Pupillo}},\ and\ \bibinfo
  {author} {\bibfnamefont {P.}~\bibnamefont {Zoller}},\ }\bibfield  {title}
  {\bibinfo {title} {Engineered open systems and quantum simulations with atoms
  and ions},\ }\href {https://link.aps.org/doi/10.1103/PhysRevA.86.023604}
  {\bibfield  {journal} {\bibinfo  {journal} {Advances in Atomic, Molecular,
  and Optical Physics}\ }\textbf {\bibinfo {volume} {61}},\ \bibinfo {pages}
  {1} (\bibinfo {year} {2012})}\BibitemShut {NoStop}%
\bibitem [{\citenamefont {Caballar}\ \emph {et~al.}(2014)\citenamefont
  {Caballar}, \citenamefont {Diehl}, \citenamefont {M\"akel\"a}, \citenamefont
  {Oberthaler},\ and\ \citenamefont {Watanabe}}]{CaballarWatanabe2014}%
  \BibitemOpen
  \bibfield  {author} {\bibinfo {author} {\bibfnamefont {R.~C.~F.}\
  \bibnamefont {Caballar}}, \bibinfo {author} {\bibfnamefont {S.}~\bibnamefont
  {Diehl}}, \bibinfo {author} {\bibfnamefont {H.}~\bibnamefont {M\"akel\"a}},
  \bibinfo {author} {\bibfnamefont {M.}~\bibnamefont {Oberthaler}},\ and\
  \bibinfo {author} {\bibfnamefont {G.}~\bibnamefont {Watanabe}},\ }\bibfield
  {title} {\bibinfo {title} {Dissipative preparation of phase- and
  number-squeezed states with ultracold atoms},\ }\href
  {https://doi.org/10.1103/PhysRevA.89.013620} {\bibfield  {journal} {\bibinfo
  {journal} {Phys. Rev. A}\ }\textbf {\bibinfo {volume} {89}},\ \bibinfo
  {pages} {013620} (\bibinfo {year} {2014})}\BibitemShut {NoStop}%
\bibitem [{\citenamefont {Syassen}\ \emph
  {et~al.}(2008{\natexlab{a}})\citenamefont {Syassen}, \citenamefont {Bauer},
  \citenamefont {Lettner}, \citenamefont {Volz}, \citenamefont {Dietze},
  \citenamefont {Garcia-Ripoll}, \citenamefont {Cirac}, \citenamefont {Rempe},\
  and\ \citenamefont {D\"urr}}]{SyassenDuerr2008}%
  \BibitemOpen
  \bibfield  {author} {\bibinfo {author} {\bibfnamefont {N.}~\bibnamefont
  {Syassen}}, \bibinfo {author} {\bibfnamefont {D.~M.}\ \bibnamefont {Bauer}},
  \bibinfo {author} {\bibfnamefont {M.}~\bibnamefont {Lettner}}, \bibinfo
  {author} {\bibfnamefont {T.}~\bibnamefont {Volz}}, \bibinfo {author}
  {\bibfnamefont {D.}~\bibnamefont {Dietze}}, \bibinfo {author} {\bibfnamefont
  {J.~J.}\ \bibnamefont {Garcia-Ripoll}}, \bibinfo {author} {\bibfnamefont
  {J.~I.}\ \bibnamefont {Cirac}}, \bibinfo {author} {\bibfnamefont
  {G.}~\bibnamefont {Rempe}},\ and\ \bibinfo {author} {\bibfnamefont
  {S.}~\bibnamefont {D\"urr}},\ }\bibfield  {title} {\bibinfo {title} {Strong
  dissipation inhibits losses and induces correlations in cold molecular
  gases},\ }\href@noop {} {\bibfield  {journal} {\bibinfo  {journal} {Science}\
  }\textbf {\bibinfo {volume} {320}},\ \bibinfo {pages} {1329} (\bibinfo {year}
  {2008}{\natexlab{a}})}\BibitemShut {NoStop}%
\bibitem [{\citenamefont {Barreiro}\ \emph {et~al.}(2011)\citenamefont
  {Barreiro}, \citenamefont {Müller}, \citenamefont {Schindler}, \citenamefont
  {Nigg}, \citenamefont {Monz}, \citenamefont {Chwalla}, \citenamefont
  {Hennrich}, \citenamefont {Roos}, \citenamefont {Zoller},\ and\ \citenamefont
  {Blatt}}]{BarreiroBlatt2011}%
  \BibitemOpen
  \bibfield  {author} {\bibinfo {author} {\bibfnamefont {J.~T.}\ \bibnamefont
  {Barreiro}}, \bibinfo {author} {\bibfnamefont {M.}~\bibnamefont {Müller}},
  \bibinfo {author} {\bibfnamefont {P.}~\bibnamefont {Schindler}}, \bibinfo
  {author} {\bibfnamefont {D.}~\bibnamefont {Nigg}}, \bibinfo {author}
  {\bibfnamefont {T.}~\bibnamefont {Monz}}, \bibinfo {author} {\bibfnamefont
  {M.}~\bibnamefont {Chwalla}}, \bibinfo {author} {\bibfnamefont
  {M.}~\bibnamefont {Hennrich}}, \bibinfo {author} {\bibfnamefont {C.~F.}\
  \bibnamefont {Roos}}, \bibinfo {author} {\bibfnamefont {P.}~\bibnamefont
  {Zoller}},\ and\ \bibinfo {author} {\bibfnamefont {R.}~\bibnamefont
  {Blatt}},\ }\bibfield  {title} {\bibinfo {title} {An open-system quantum
  simulator with trapped ions},\ }\href@noop {} {\bibfield  {journal} {\bibinfo
   {journal} {Nature}\ }\textbf {\bibinfo {volume} {470}},\ \bibinfo {pages}
  {486} (\bibinfo {year} {2011})}\BibitemShut {NoStop}%
\bibitem [{\citenamefont {Viciani}\ \emph {et~al.}(2015)\citenamefont
  {Viciani}, \citenamefont {Lima}, \citenamefont {Bellini},\ and\ \citenamefont
  {Caruso}}]{VicianiCaruso2015}%
  \BibitemOpen
  \bibfield  {author} {\bibinfo {author} {\bibfnamefont {S.}~\bibnamefont
  {Viciani}}, \bibinfo {author} {\bibfnamefont {M.}~\bibnamefont {Lima}},
  \bibinfo {author} {\bibfnamefont {M.}~\bibnamefont {Bellini}},\ and\ \bibinfo
  {author} {\bibfnamefont {F.}~\bibnamefont {Caruso}},\ }\bibfield  {title}
  {\bibinfo {title} {Observation of noise-assisted transport in an all-optical
  cavity-based network},\ }\href
  {https://doi.org/10.1103/PhysRevLett.115.083601} {\bibfield  {journal}
  {\bibinfo  {journal} {Phys. Rev. Lett.}\ }\textbf {\bibinfo {volume} {115}},\
  \bibinfo {pages} {083601} (\bibinfo {year} {2015})}\BibitemShut {NoStop}%
\bibitem [{\citenamefont {Maier}\ \emph {et~al.}(2019)\citenamefont {Maier},
  \citenamefont {Brydges}, \citenamefont {Jurcevic}, \citenamefont {Trautmann},
  \citenamefont {Hempel}, \citenamefont {Lanyon}, \citenamefont {Hauke},
  \citenamefont {Blatt},\ and\ \citenamefont {Roos}}]{MaierRoos2019}%
  \BibitemOpen
  \bibfield  {author} {\bibinfo {author} {\bibfnamefont {C.}~\bibnamefont
  {Maier}}, \bibinfo {author} {\bibfnamefont {T.}~\bibnamefont {Brydges}},
  \bibinfo {author} {\bibfnamefont {P.}~\bibnamefont {Jurcevic}}, \bibinfo
  {author} {\bibfnamefont {N.}~\bibnamefont {Trautmann}}, \bibinfo {author}
  {\bibfnamefont {C.}~\bibnamefont {Hempel}}, \bibinfo {author} {\bibfnamefont
  {B.~P.}\ \bibnamefont {Lanyon}}, \bibinfo {author} {\bibfnamefont
  {P.}~\bibnamefont {Hauke}}, \bibinfo {author} {\bibfnamefont
  {R.}~\bibnamefont {Blatt}},\ and\ \bibinfo {author} {\bibfnamefont {C.~F.}\
  \bibnamefont {Roos}},\ }\bibfield  {title} {\bibinfo {title}
  {Environment-assisted quantum transport in a 10-qubit network},\ }\href
  {https://doi.org/10.1103/PhysRevLett.122.050501} {\bibfield  {journal}
  {\bibinfo  {journal} {Phys. Rev. Lett.}\ }\textbf {\bibinfo {volume} {122}},\
  \bibinfo {pages} {050501} (\bibinfo {year} {2019})}\BibitemShut {NoStop}%
\bibitem [{\citenamefont {Dolgirev}\ \emph {et~al.}(2020)\citenamefont
  {Dolgirev}, \citenamefont {Marino}, \citenamefont {Sels},\ and\ \citenamefont
  {Demler}}]{dolgirev_impurity_noise}%
  \BibitemOpen
  \bibfield  {author} {\bibinfo {author} {\bibfnamefont {P.~E.}\ \bibnamefont
  {Dolgirev}}, \bibinfo {author} {\bibfnamefont {J.}~\bibnamefont {Marino}},
  \bibinfo {author} {\bibfnamefont {D.}~\bibnamefont {Sels}},\ and\ \bibinfo
  {author} {\bibfnamefont {E.}~\bibnamefont {Demler}},\ }\bibfield  {title}
  {\bibinfo {title} {Non-gaussian correlations imprinted by local dephasing in
  fermionic wires},\ }\href {https://doi.org/10.1103/PhysRevB.102.100301}
  {\bibfield  {journal} {\bibinfo  {journal} {Phys. Rev. B}\ }\textbf {\bibinfo
  {volume} {102}},\ \bibinfo {pages} {100301} (\bibinfo {year}
  {2020})}\BibitemShut {NoStop}%
\bibitem [{\citenamefont {Barontini}\ \emph {et~al.}(2013)\citenamefont
  {Barontini}, \citenamefont {Labouvie}, \citenamefont {Stubenrauch},
  \citenamefont {Vogler}, \citenamefont {Guarrera},\ and\ \citenamefont
  {Ott}}]{BarontiniOtt2013}%
  \BibitemOpen
  \bibfield  {author} {\bibinfo {author} {\bibfnamefont {G.}~\bibnamefont
  {Barontini}}, \bibinfo {author} {\bibfnamefont {R.}~\bibnamefont {Labouvie}},
  \bibinfo {author} {\bibfnamefont {F.}~\bibnamefont {Stubenrauch}}, \bibinfo
  {author} {\bibfnamefont {A.}~\bibnamefont {Vogler}}, \bibinfo {author}
  {\bibfnamefont {V.}~\bibnamefont {Guarrera}},\ and\ \bibinfo {author}
  {\bibfnamefont {H.}~\bibnamefont {Ott}},\ }\bibfield  {title} {\bibinfo
  {title} {Controlling the dynamics of an open many-body quantum system with
  localized dissipation},\ }\href
  {https://doi.org/10.1103/PhysRevLett.110.035302} {\bibfield  {journal}
  {\bibinfo  {journal} {Phys. Rev. Lett.}\ }\textbf {\bibinfo {volume} {110}},\
  \bibinfo {pages} {035302} (\bibinfo {year} {2013})}\BibitemShut {NoStop}%
\bibitem [{\citenamefont {Labouvie}\ \emph {et~al.}(2016)\citenamefont
  {Labouvie}, \citenamefont {Santra}, \citenamefont {Heun},\ and\ \citenamefont
  {Ott}}]{LabouvieOtt2016}%
  \BibitemOpen
  \bibfield  {author} {\bibinfo {author} {\bibfnamefont {R.}~\bibnamefont
  {Labouvie}}, \bibinfo {author} {\bibfnamefont {B.}~\bibnamefont {Santra}},
  \bibinfo {author} {\bibfnamefont {S.}~\bibnamefont {Heun}},\ and\ \bibinfo
  {author} {\bibfnamefont {H.}~\bibnamefont {Ott}},\ }\bibfield  {title}
  {\bibinfo {title} {Bistability in a driven-dissipative superfluid},\ }\href
  {https://doi.org/10.1103/PhysRevLett.116.235302} {\bibfield  {journal}
  {\bibinfo  {journal} {Phys. Rev. Lett.}\ }\textbf {\bibinfo {volume} {116}},\
  \bibinfo {pages} {235302} (\bibinfo {year} {2016})}\BibitemShut {NoStop}%
\bibitem [{\citenamefont {Corman}\ \emph {et~al.}(2019)\citenamefont {Corman},
  \citenamefont {Fabritius}, \citenamefont {H\"ausler}, \citenamefont {Mohan},
  \citenamefont {Dogra}, \citenamefont {Husmann}, \citenamefont {Lebrat},\ and\
  \citenamefont {Esslinger}}]{CormanEsslinger2019}%
  \BibitemOpen
  \bibfield  {author} {\bibinfo {author} {\bibfnamefont {L.}~\bibnamefont
  {Corman}}, \bibinfo {author} {\bibfnamefont {P.}~\bibnamefont {Fabritius}},
  \bibinfo {author} {\bibfnamefont {S.}~\bibnamefont {H\"ausler}}, \bibinfo
  {author} {\bibfnamefont {J.}~\bibnamefont {Mohan}}, \bibinfo {author}
  {\bibfnamefont {L.~H.}\ \bibnamefont {Dogra}}, \bibinfo {author}
  {\bibfnamefont {D.}~\bibnamefont {Husmann}}, \bibinfo {author} {\bibfnamefont
  {M.}~\bibnamefont {Lebrat}},\ and\ \bibinfo {author} {\bibfnamefont
  {T.}~\bibnamefont {Esslinger}},\ }\bibfield  {title} {\bibinfo {title}
  {Quantized conductance through a dissipative atomic point contact},\ }\href
  {https://doi.org/10.1103/PhysRevA.100.053605} {\bibfield  {journal} {\bibinfo
   {journal} {Phys. Rev. A}\ }\textbf {\bibinfo {volume} {100}},\ \bibinfo
  {pages} {053605} (\bibinfo {year} {2019})}\BibitemShut {NoStop}%
\bibitem [{\citenamefont {Lebrat}\ \emph {et~al.}(2019)\citenamefont {Lebrat},
  \citenamefont {H\"ausler}, \citenamefont {Fabritius}, \citenamefont
  {Husmann}, \citenamefont {Corman},\ and\ \citenamefont
  {Esslinger}}]{LebratEsslinger2019}%
  \BibitemOpen
  \bibfield  {author} {\bibinfo {author} {\bibfnamefont {M.}~\bibnamefont
  {Lebrat}}, \bibinfo {author} {\bibfnamefont {S.}~\bibnamefont {H\"ausler}},
  \bibinfo {author} {\bibfnamefont {P.}~\bibnamefont {Fabritius}}, \bibinfo
  {author} {\bibfnamefont {D.}~\bibnamefont {Husmann}}, \bibinfo {author}
  {\bibfnamefont {L.}~\bibnamefont {Corman}},\ and\ \bibinfo {author}
  {\bibfnamefont {T.}~\bibnamefont {Esslinger}},\ }\bibfield  {title} {\bibinfo
  {title} {Quantized conductance through a spin-selective atomic point
  contact},\ }\href {https://doi.org/10.1103/PhysRevLett.123.193605} {\bibfield
   {journal} {\bibinfo  {journal} {Phys. Rev. Lett.}\ }\textbf {\bibinfo
  {volume} {123}},\ \bibinfo {pages} {193605} (\bibinfo {year}
  {2019})}\BibitemShut {NoStop}%
\bibitem [{\citenamefont {Mivehvar}\ \emph {et~al.}(2021)\citenamefont
  {Mivehvar}, \citenamefont {Piazza}, \citenamefont {Donner},\ and\
  \citenamefont {Ritsch}}]{mivehvar_cavity_review}%
  \BibitemOpen
  \bibfield  {author} {\bibinfo {author} {\bibfnamefont {F.}~\bibnamefont
  {Mivehvar}}, \bibinfo {author} {\bibfnamefont {F.}~\bibnamefont {Piazza}},
  \bibinfo {author} {\bibfnamefont {T.}~\bibnamefont {Donner}},\ and\ \bibinfo
  {author} {\bibfnamefont {H.}~\bibnamefont {Ritsch}},\ }\bibfield  {title}
  {\bibinfo {title} {Cavity qed with quantum gases: new paradigms in many-body
  physics},\ }\href {https://doi.org/10.1080/00018732.2021.1969727} {\bibfield
  {journal} {\bibinfo  {journal} {Advances in Physics}\ }\textbf {\bibinfo
  {volume} {70}},\ \bibinfo {pages} {1} (\bibinfo {year} {2021})},\ \Eprint
  {https://arxiv.org/abs/https://doi.org/10.1080/00018732.2021.1969727}
  {https://doi.org/10.1080/00018732.2021.1969727} \BibitemShut {NoStop}%
\bibitem [{\citenamefont {Jones}\ \emph {et~al.}(2006)\citenamefont {Jones},
  \citenamefont {Tiesinga}, \citenamefont {Lett},\ and\ \citenamefont
  {Julienne}}]{jones_photoassociation_review}%
  \BibitemOpen
  \bibfield  {author} {\bibinfo {author} {\bibfnamefont {K.~M.}\ \bibnamefont
  {Jones}}, \bibinfo {author} {\bibfnamefont {E.}~\bibnamefont {Tiesinga}},
  \bibinfo {author} {\bibfnamefont {P.~D.}\ \bibnamefont {Lett}},\ and\
  \bibinfo {author} {\bibfnamefont {P.~S.}\ \bibnamefont {Julienne}},\
  }\bibfield  {title} {\bibinfo {title} {Ultracold photoassociation
  spectroscopy: Long-range molecules and atomic scattering},\ }\href
  {https://doi.org/10.1103/RevModPhys.78.483} {\bibfield  {journal} {\bibinfo
  {journal} {Rev. Mod. Phys.}\ }\textbf {\bibinfo {volume} {78}},\ \bibinfo
  {pages} {483} (\bibinfo {year} {2006})}\BibitemShut {NoStop}%
\bibitem [{\citenamefont {Konishi}\ \emph {et~al.}(2021)\citenamefont
  {Konishi}, \citenamefont {Roux}, \citenamefont {Helson},\ and\ \citenamefont
  {Brantut}}]{konishi_polariton_cavity}%
  \BibitemOpen
  \bibfield  {author} {\bibinfo {author} {\bibfnamefont {H.}~\bibnamefont
  {Konishi}}, \bibinfo {author} {\bibfnamefont {K.}~\bibnamefont {Roux}},
  \bibinfo {author} {\bibfnamefont {V.}~\bibnamefont {Helson}},\ and\ \bibinfo
  {author} {\bibfnamefont {J.-P.}\ \bibnamefont {Brantut}},\ }\bibfield
  {title} {\bibinfo {title} {Universal pair polaritons in a strongly
  interacting fermi gas},\ }\href {https://doi.org/10.1038/s41586-021-03731-9}
  {\bibfield  {journal} {\bibinfo  {journal} {Nature}\ }\textbf {\bibinfo
  {volume} {596}},\ \bibinfo {pages} {509} (\bibinfo {year}
  {2021})}\BibitemShut {NoStop}%
\bibitem [{\citenamefont {Yamamoto}\ \emph {et~al.}(2019)\citenamefont
  {Yamamoto}, \citenamefont {Nakagawa}, \citenamefont {Adachi}, \citenamefont
  {Takasan}, \citenamefont {Ueda},\ and\ \citenamefont
  {Kawakami}}]{Yamamoto_PhysRevLett.123.123601}%
  \BibitemOpen
  \bibfield  {author} {\bibinfo {author} {\bibfnamefont {K.}~\bibnamefont
  {Yamamoto}}, \bibinfo {author} {\bibfnamefont {M.}~\bibnamefont {Nakagawa}},
  \bibinfo {author} {\bibfnamefont {K.}~\bibnamefont {Adachi}}, \bibinfo
  {author} {\bibfnamefont {K.}~\bibnamefont {Takasan}}, \bibinfo {author}
  {\bibfnamefont {M.}~\bibnamefont {Ueda}},\ and\ \bibinfo {author}
  {\bibfnamefont {N.}~\bibnamefont {Kawakami}},\ }\bibfield  {title} {\bibinfo
  {title} {Theory of non-hermitian fermionic superfluidity with a
  complex-valued interaction},\ }\href
  {https://doi.org/10.1103/PhysRevLett.123.123601} {\bibfield  {journal}
  {\bibinfo  {journal} {Phys. Rev. Lett.}\ }\textbf {\bibinfo {volume} {123}},\
  \bibinfo {pages} {123601} (\bibinfo {year} {2019})}\BibitemShut {NoStop}%
\bibitem [{\citenamefont {Nakagawa}\ \emph {et~al.}(2020)\citenamefont
  {Nakagawa}, \citenamefont {Tsuji}, \citenamefont {Kawakami},\ and\
  \citenamefont {Ueda}}]{Nakagawa2_PhysRevLett.124.147203}%
  \BibitemOpen
  \bibfield  {author} {\bibinfo {author} {\bibfnamefont {M.}~\bibnamefont
  {Nakagawa}}, \bibinfo {author} {\bibfnamefont {N.}~\bibnamefont {Tsuji}},
  \bibinfo {author} {\bibfnamefont {N.}~\bibnamefont {Kawakami}},\ and\
  \bibinfo {author} {\bibfnamefont {M.}~\bibnamefont {Ueda}},\ }\bibfield
  {title} {\bibinfo {title} {Dynamical sign reversal of magnetic correlations
  in dissipative hubbard models},\ }\href
  {https://doi.org/10.1103/PhysRevLett.124.147203} {\bibfield  {journal}
  {\bibinfo  {journal} {Phys. Rev. Lett.}\ }\textbf {\bibinfo {volume} {124}},\
  \bibinfo {pages} {147203} (\bibinfo {year} {2020})}\BibitemShut {NoStop}%
\bibitem [{\citenamefont {Yamamoto}\ \emph {et~al.}(2022)\citenamefont
  {Yamamoto}, \citenamefont {Nakagawa}, \citenamefont {Tezuka}, \citenamefont
  {Ueda},\ and\ \citenamefont {Kawakami}}]{Yamamoto2_PhysRevB.105.205125}%
  \BibitemOpen
  \bibfield  {author} {\bibinfo {author} {\bibfnamefont {K.}~\bibnamefont
  {Yamamoto}}, \bibinfo {author} {\bibfnamefont {M.}~\bibnamefont {Nakagawa}},
  \bibinfo {author} {\bibfnamefont {M.}~\bibnamefont {Tezuka}}, \bibinfo
  {author} {\bibfnamefont {M.}~\bibnamefont {Ueda}},\ and\ \bibinfo {author}
  {\bibfnamefont {N.}~\bibnamefont {Kawakami}},\ }\bibfield  {title} {\bibinfo
  {title} {Universal properties of dissipative tomonaga-luttinger liquids: Case
  study of a non-hermitian xxz spin chain},\ }\href
  {https://doi.org/10.1103/PhysRevB.105.205125} {\bibfield  {journal} {\bibinfo
   {journal} {Phys. Rev. B}\ }\textbf {\bibinfo {volume} {105}},\ \bibinfo
  {pages} {205125} (\bibinfo {year} {2022})}\BibitemShut {NoStop}%
\bibitem [{\citenamefont {Ashida}\ \emph
  {et~al.}(2020{\natexlab{b}})\citenamefont {Ashida}, \citenamefont {Gong},\
  and\ \citenamefont {Ueda}}]{Ashida_AdvanceinPhys_2020}%
  \BibitemOpen
  \bibfield  {author} {\bibinfo {author} {\bibfnamefont {Y.}~\bibnamefont
  {Ashida}}, \bibinfo {author} {\bibfnamefont {Z.}~\bibnamefont {Gong}},\ and\
  \bibinfo {author} {\bibfnamefont {M.}~\bibnamefont {Ueda}},\ }\bibfield
  {title} {\bibinfo {title} {Non-hermitian physics},\ }\href
  {https://doi.org/10.1080/00018732.2021.1876991} {\bibfield  {journal}
  {\bibinfo  {journal} {Advances in Physics}\ }\textbf {\bibinfo {volume}
  {69}},\ \bibinfo {pages} {249} (\bibinfo {year} {2020}{\natexlab{b}})},\
  \Eprint {https://arxiv.org/abs/https://doi.org/10.1080/00018732.2021.1876991}
  {https://doi.org/10.1080/00018732.2021.1876991} \BibitemShut {NoStop}%
\bibitem [{\citenamefont {Nakagawa}\ \emph {et~al.}(2018)\citenamefont
  {Nakagawa}, \citenamefont {Kawakami},\ and\ \citenamefont
  {Ueda}}]{Nakagawa_PhysRevLett.121.203001}%
  \BibitemOpen
  \bibfield  {author} {\bibinfo {author} {\bibfnamefont {M.}~\bibnamefont
  {Nakagawa}}, \bibinfo {author} {\bibfnamefont {N.}~\bibnamefont {Kawakami}},\
  and\ \bibinfo {author} {\bibfnamefont {M.}~\bibnamefont {Ueda}},\ }\bibfield
  {title} {\bibinfo {title} {Non-hermitian kondo effect in ultracold
  alkaline-earth atoms},\ }\href
  {https://doi.org/10.1103/PhysRevLett.121.203001} {\bibfield  {journal}
  {\bibinfo  {journal} {Phys. Rev. Lett.}\ }\textbf {\bibinfo {volume} {121}},\
  \bibinfo {pages} {203001} (\bibinfo {year} {2018})}\BibitemShut {NoStop}%
\bibitem [{\citenamefont {Yamamoto}\ and\ \citenamefont
  {Kawakami}(2023)}]{Yamamoto3_PhysRevB.107.045110}%
  \BibitemOpen
  \bibfield  {author} {\bibinfo {author} {\bibfnamefont {K.}~\bibnamefont
  {Yamamoto}}\ and\ \bibinfo {author} {\bibfnamefont {N.}~\bibnamefont
  {Kawakami}},\ }\bibfield  {title} {\bibinfo {title} {Universal description of
  dissipative tomonaga-luttinger liquids with $\mathrm{SU}(n)$ spin symmetry:
  Exact spectrum and critical exponents},\ }\href
  {https://doi.org/10.1103/PhysRevB.107.045110} {\bibfield  {journal} {\bibinfo
   {journal} {Phys. Rev. B}\ }\textbf {\bibinfo {volume} {107}},\ \bibinfo
  {pages} {045110} (\bibinfo {year} {2023})}\BibitemShut {NoStop}%
\bibitem [{\citenamefont {Han}\ \emph {et~al.}(2023)\citenamefont {Han},
  \citenamefont {Schultz},\ and\ \citenamefont {Kim}}]{han2023complex}%
  \BibitemOpen
  \bibfield  {author} {\bibinfo {author} {\bibfnamefont {S.}~\bibnamefont
  {Han}}, \bibinfo {author} {\bibfnamefont {D.~J.}\ \bibnamefont {Schultz}},\
  and\ \bibinfo {author} {\bibfnamefont {Y.~B.}\ \bibnamefont {Kim}},\
  }\href@noop {} {\bibinfo {title} {Complex fixed points of the non-hermitian
  kondo model in a luttinger liquid}} (\bibinfo {year} {2023}),\ \Eprint
  {https://arxiv.org/abs/2302.07883} {arXiv:2302.07883 [cond-mat.str-el]}
  \BibitemShut {NoStop}%
\bibitem [{\citenamefont {Damanet}\ \emph
  {et~al.}(2019{\natexlab{a}})\citenamefont {Damanet}, \citenamefont
  {Mascarenhas}, \citenamefont {Pekker},\ and\ \citenamefont
  {Daley}}]{damanet_controlling_2019}%
  \BibitemOpen
  \bibfield  {author} {\bibinfo {author} {\bibfnamefont {F.}~\bibnamefont
  {Damanet}}, \bibinfo {author} {\bibfnamefont {E.}~\bibnamefont
  {Mascarenhas}}, \bibinfo {author} {\bibfnamefont {D.}~\bibnamefont
  {Pekker}},\ and\ \bibinfo {author} {\bibfnamefont {A.~J.}\ \bibnamefont
  {Daley}},\ }\bibfield  {title} {\bibinfo {title} {Controlling {Quantum}
  {Transport} via {Dissipation} {Engineering}},\ }\href
  {https://doi.org/10.1103/PhysRevLett.123.180402} {\bibfield  {journal}
  {\bibinfo  {journal} {Physical Review Letters}\ }\textbf {\bibinfo {volume}
  {123}},\ \bibinfo {pages} {180402} (\bibinfo {year}
  {2019}{\natexlab{a}})}\BibitemShut {NoStop}%
\bibitem [{\citenamefont {Damanet}\ \emph
  {et~al.}(2019{\natexlab{b}})\citenamefont {Damanet}, \citenamefont
  {Mascarenhas}, \citenamefont {Pekker},\ and\ \citenamefont
  {Daley}}]{damanet_reservoir_2019a}%
  \BibitemOpen
  \bibfield  {author} {\bibinfo {author} {\bibfnamefont {F.}~\bibnamefont
  {Damanet}}, \bibinfo {author} {\bibfnamefont {E.}~\bibnamefont
  {Mascarenhas}}, \bibinfo {author} {\bibfnamefont {D.}~\bibnamefont
  {Pekker}},\ and\ \bibinfo {author} {\bibfnamefont {A.~J.}\ \bibnamefont
  {Daley}},\ }\bibfield  {title} {\bibinfo {title} {Reservoir engineering of
  {C}ooper-pair-assisted transport with cold atoms},\ }\href
  {https://doi.org/10.1088/1367-2630/ab4f5d} {\bibfield  {journal} {\bibinfo
  {journal} {New Journal of Physics}\ }\textbf {\bibinfo {volume} {21}},\
  \bibinfo {pages} {115001} (\bibinfo {year} {2019}{\natexlab{b}})}\BibitemShut
  {NoStop}%
\bibitem [{\citenamefont {Damanet}\ \emph
  {et~al.}(2019{\natexlab{c}})\citenamefont {Damanet}, \citenamefont
  {Mascarenhas}, \citenamefont {Pekker},\ and\ \citenamefont
  {Daley}}]{damanet_reservoir_2019b}%
  \BibitemOpen
  \bibfield  {author} {\bibinfo {author} {\bibfnamefont {F.}~\bibnamefont
  {Damanet}}, \bibinfo {author} {\bibfnamefont {E.}~\bibnamefont
  {Mascarenhas}}, \bibinfo {author} {\bibfnamefont {D.}~\bibnamefont
  {Pekker}},\ and\ \bibinfo {author} {\bibfnamefont {A.~J.}\ \bibnamefont
  {Daley}},\ }\bibfield  {title} {\bibinfo {title} {Reservoir engineering of
  {Cooper}-pair-assisted transport with cold atoms},\ }\href
  {https://doi.org/10.1088/1367-2630/ab4f5d} {\bibfield  {journal} {\bibinfo
  {journal} {New Journal of Physics}\ }\textbf {\bibinfo {volume} {21}},\
  \bibinfo {pages} {115001} (\bibinfo {year} {2019}{\natexlab{c}})}\BibitemShut
  {NoStop}%
\bibitem [{\citenamefont {Yamamoto}\ \emph {et~al.}(2021)\citenamefont
  {Yamamoto}, \citenamefont {Nakagawa}, \citenamefont {Tsuji}, \citenamefont
  {Ueda},\ and\ \citenamefont {Kawakami}}]{YamamotoKawakami2021}%
  \BibitemOpen
  \bibfield  {author} {\bibinfo {author} {\bibfnamefont {K.}~\bibnamefont
  {Yamamoto}}, \bibinfo {author} {\bibfnamefont {M.}~\bibnamefont {Nakagawa}},
  \bibinfo {author} {\bibfnamefont {N.}~\bibnamefont {Tsuji}}, \bibinfo
  {author} {\bibfnamefont {M.}~\bibnamefont {Ueda}},\ and\ \bibinfo {author}
  {\bibfnamefont {N.}~\bibnamefont {Kawakami}},\ }\bibfield  {title} {\bibinfo
  {title} {Collective {E}xcitations and {N}onequilibrium {P}hase {T}ransition
  in {D}issipative {F}ermionic {S}uperfluids},\ }\href
  {https://doi.org/10.1103/PhysRevLett.127.055301} {\bibfield  {journal}
  {\bibinfo  {journal} {Phys. Rev. Lett.}\ }\textbf {\bibinfo {volume} {127}},\
  \bibinfo {pages} {055301} (\bibinfo {year} {2021})}\BibitemShut {NoStop}%
\bibitem [{\citenamefont {Mazza}\ and\ \citenamefont
  {Schirò}(2023)}]{MazzaSchiro2023}%
  \BibitemOpen
  \bibfield  {author} {\bibinfo {author} {\bibfnamefont {G.}~\bibnamefont
  {Mazza}}\ and\ \bibinfo {author} {\bibfnamefont {M.}~\bibnamefont
  {Schirò}},\ }\href@noop {} {\bibinfo {title} {Dissipative dynamics of a
  fermionic superfluid with two-body losses}} (\bibinfo {year} {2023}),\
  \Eprint {https://arxiv.org/abs/2208.02739} {arXiv:2208.02739
  [cond-mat.str-el]} \BibitemShut {NoStop}%
\bibitem [{\citenamefont {Auletta}\ \emph {et~al.}(2009)\citenamefont
  {Auletta}, \citenamefont {Fortunato},\ and\ \citenamefont
  {Parisi}}]{auletta_fortunato_parisi_2009}%
  \BibitemOpen
  \bibfield  {author} {\bibinfo {author} {\bibfnamefont {G.}~\bibnamefont
  {Auletta}}, \bibinfo {author} {\bibfnamefont {M.}~\bibnamefont {Fortunato}},\
  and\ \bibinfo {author} {\bibfnamefont {G.}~\bibnamefont {Parisi}},\ }\href
  {https://doi.org/10.1017/CBO9780511813955} {\emph {\bibinfo {title} {Quantum
  Mechanics}}}\ (\bibinfo  {publisher} {Cambridge University Press},\ \bibinfo
  {year} {2009})\BibitemShut {NoStop}%
\bibitem [{\citenamefont {Kamenev}(2011)}]{kamenev_2011}%
  \BibitemOpen
  \bibfield  {author} {\bibinfo {author} {\bibfnamefont {A.}~\bibnamefont
  {Kamenev}},\ }\href {https://doi.org/10.1017/CBO9781139003667} {\emph
  {\bibinfo {title} {Field Theory of Non-Equilibrium Systems}}}\ (\bibinfo
  {publisher} {Cambridge University Press},\ \bibinfo {year}
  {2011})\BibitemShut {NoStop}%
\bibitem [{\citenamefont {Sieberer}\ \emph
  {et~al.}(2016{\natexlab{a}})\citenamefont {Sieberer}, \citenamefont
  {Buchhold},\ and\ \citenamefont {Diehl}}]{Sieberer_IOP_2016}%
  \BibitemOpen
  \bibfield  {author} {\bibinfo {author} {\bibfnamefont {L.~M.}\ \bibnamefont
  {Sieberer}}, \bibinfo {author} {\bibfnamefont {M.}~\bibnamefont {Buchhold}},\
  and\ \bibinfo {author} {\bibfnamefont {S.}~\bibnamefont {Diehl}},\ }\bibfield
   {title} {\bibinfo {title} {Keldysh field theory for driven open quantum
  systems},\ }\href {https://doi.org/10.1088/0034-4885/79/9/096001} {\bibfield
  {journal} {\bibinfo  {journal} {Reports on Progress in Physics}\ }\textbf
  {\bibinfo {volume} {79}},\ \bibinfo {pages} {096001} (\bibinfo {year}
  {2016}{\natexlab{a}})}\BibitemShut {NoStop}%
\bibitem [{\citenamefont {Feynman}\ and\ \citenamefont
  {Vernon}(1963)}]{FEYNMAN1963118}%
  \BibitemOpen
  \bibfield  {author} {\bibinfo {author} {\bibfnamefont {R.}~\bibnamefont
  {Feynman}}\ and\ \bibinfo {author} {\bibfnamefont {F.}~\bibnamefont
  {Vernon}},\ }\bibfield  {title} {\bibinfo {title} {The theory of a general
  quantum system interacting with a linear dissipative system},\ }\href
  {https://doi.org/https://doi.org/10.1016/0003-4916(63)90068-X} {\bibfield
  {journal} {\bibinfo  {journal} {Annals of Physics}\ }\textbf {\bibinfo
  {volume} {24}},\ \bibinfo {pages} {118} (\bibinfo {year} {1963})}\BibitemShut
  {NoStop}%
\bibitem [{\citenamefont {Sols}\ and\ \citenamefont
  {Guinea}(1987)}]{Sols_PhysRevB.36.7775}%
  \BibitemOpen
  \bibfield  {author} {\bibinfo {author} {\bibfnamefont {F.}~\bibnamefont
  {Sols}}\ and\ \bibinfo {author} {\bibfnamefont {F.}~\bibnamefont {Guinea}},\
  }\bibfield  {title} {\bibinfo {title} {Bulk and surface diffusion of heavy
  particles in metals: A path-integral approach},\ }\href
  {https://doi.org/10.1103/PhysRevB.36.7775} {\bibfield  {journal} {\bibinfo
  {journal} {Phys. Rev. B}\ }\textbf {\bibinfo {volume} {36}},\ \bibinfo
  {pages} {7775} (\bibinfo {year} {1987})}\BibitemShut {NoStop}%
\bibitem [{\citenamefont {Cazalilla}\ \emph {et~al.}(2006)\citenamefont
  {Cazalilla}, \citenamefont {Sols},\ and\ \citenamefont
  {Guinea}}]{Cazalilla_PhysRevLett.97.076401}%
  \BibitemOpen
  \bibfield  {author} {\bibinfo {author} {\bibfnamefont {M.~A.}\ \bibnamefont
  {Cazalilla}}, \bibinfo {author} {\bibfnamefont {F.}~\bibnamefont {Sols}},\
  and\ \bibinfo {author} {\bibfnamefont {F.}~\bibnamefont {Guinea}},\
  }\bibfield  {title} {\bibinfo {title} {Dissipation-driven quantum phase
  transitions in a tomonaga-luttinger liquid electrostatically coupled to a
  metallic gate},\ }\href {https://doi.org/10.1103/PhysRevLett.97.076401}
  {\bibfield  {journal} {\bibinfo  {journal} {Phys. Rev. Lett.}\ }\textbf
  {\bibinfo {volume} {97}},\ \bibinfo {pages} {076401} (\bibinfo {year}
  {2006})}\BibitemShut {NoStop}%
\bibitem [{\citenamefont {Lobos}\ and\ \citenamefont
  {Giamarchi}(2010)}]{Lobos_PhysRevB.82.104517}%
  \BibitemOpen
  \bibfield  {author} {\bibinfo {author} {\bibfnamefont {A.~M.}\ \bibnamefont
  {Lobos}}\ and\ \bibinfo {author} {\bibfnamefont {T.}~\bibnamefont
  {Giamarchi}},\ }\bibfield  {title} {\bibinfo {title} {Dissipative phase
  fluctuations in superconducting wires capacitively coupled to diffusive
  metals},\ }\href {https://doi.org/10.1103/PhysRevB.82.104517} {\bibfield
  {journal} {\bibinfo  {journal} {Phys. Rev. B}\ }\textbf {\bibinfo {volume}
  {82}},\ \bibinfo {pages} {104517} (\bibinfo {year} {2010})}\BibitemShut
  {NoStop}%
\bibitem [{\citenamefont {Malatsetxebarria}\ \emph {et~al.}(2013)\citenamefont
  {Malatsetxebarria}, \citenamefont {Cai}, \citenamefont {Schollw\"ock},\ and\
  \citenamefont {Cazalilla}}]{Malatsetxebarria_PhysRevA.88.063630}%
  \BibitemOpen
  \bibfield  {author} {\bibinfo {author} {\bibfnamefont {E.}~\bibnamefont
  {Malatsetxebarria}}, \bibinfo {author} {\bibfnamefont {Z.}~\bibnamefont
  {Cai}}, \bibinfo {author} {\bibfnamefont {U.}~\bibnamefont {Schollw\"ock}},\
  and\ \bibinfo {author} {\bibfnamefont {M.~A.}\ \bibnamefont {Cazalilla}},\
  }\bibfield  {title} {\bibinfo {title} {Dissipative effects on the
  superfluid-to-insulator transition in mixed-dimensional optical lattices},\
  }\href {https://doi.org/10.1103/PhysRevA.88.063630} {\bibfield  {journal}
  {\bibinfo  {journal} {Phys. Rev. A}\ }\textbf {\bibinfo {volume} {88}},\
  \bibinfo {pages} {063630} (\bibinfo {year} {2013})}\BibitemShut {NoStop}%
\bibitem [{\citenamefont {Husmann}\ \emph {et~al.}(2015)\citenamefont
  {Husmann}, \citenamefont {Uchino}, \citenamefont {Krinner}, \citenamefont
  {Lebrat}, \citenamefont {Giamarchi}, \citenamefont {Esslinger},\ and\
  \citenamefont {Brantut}}]{Husmann2015}%
  \BibitemOpen
  \bibfield  {author} {\bibinfo {author} {\bibfnamefont {D.}~\bibnamefont
  {Husmann}}, \bibinfo {author} {\bibfnamefont {S.}~\bibnamefont {Uchino}},
  \bibinfo {author} {\bibfnamefont {S.}~\bibnamefont {Krinner}}, \bibinfo
  {author} {\bibfnamefont {M.}~\bibnamefont {Lebrat}}, \bibinfo {author}
  {\bibfnamefont {T.}~\bibnamefont {Giamarchi}}, \bibinfo {author}
  {\bibfnamefont {T.}~\bibnamefont {Esslinger}},\ and\ \bibinfo {author}
  {\bibfnamefont {J.}~\bibnamefont {Brantut}},\ }\bibfield  {title} {\bibinfo
  {title} {Connecting strongly correlated superfluids by a quantum point
  contact},\ }\href@noop {} {\bibfield  {journal} {\bibinfo  {journal}
  {Science}\ }\textbf {\bibinfo {volume} {350}},\ \bibinfo {pages} {1498}
  (\bibinfo {year} {2015})}\BibitemShut {NoStop}%
\bibitem [{\citenamefont {Jin}\ \emph {et~al.}(2022)\citenamefont {Jin},
  \citenamefont {Ferreira}, \citenamefont {Filippone},\ and\ \citenamefont
  {Giamarchi}}]{jin_noise_exact_1D}%
  \BibitemOpen
  \bibfield  {author} {\bibinfo {author} {\bibfnamefont {T.}~\bibnamefont
  {Jin}}, \bibinfo {author} {\bibfnamefont {J.~a.~S.}\ \bibnamefont
  {Ferreira}}, \bibinfo {author} {\bibfnamefont {M.}~\bibnamefont
  {Filippone}},\ and\ \bibinfo {author} {\bibfnamefont {T.}~\bibnamefont
  {Giamarchi}},\ }\bibfield  {title} {\bibinfo {title} {Exact description of
  quantum stochastic models as quantum resistors},\ }\href
  {https://doi.org/10.1103/PhysRevResearch.4.013109} {\bibfield  {journal}
  {\bibinfo  {journal} {Phys. Rev. Res.}\ }\textbf {\bibinfo {volume} {4}},\
  \bibinfo {pages} {013109} (\bibinfo {year} {2022})}\BibitemShut {NoStop}%
\bibitem [{\citenamefont {Fr\"oml}\ \emph {et~al.}(2020)\citenamefont
  {Fr\"oml}, \citenamefont {Muckel}, \citenamefont {Kollath}, \citenamefont
  {Chiocchetta},\ and\ \citenamefont {Diehl}}]{froeml_ultracold_2020}%
  \BibitemOpen
  \bibfield  {author} {\bibinfo {author} {\bibfnamefont {H.}~\bibnamefont
  {Fr\"oml}}, \bibinfo {author} {\bibfnamefont {C.}~\bibnamefont {Muckel}},
  \bibinfo {author} {\bibfnamefont {C.}~\bibnamefont {Kollath}}, \bibinfo
  {author} {\bibfnamefont {A.}~\bibnamefont {Chiocchetta}},\ and\ \bibinfo
  {author} {\bibfnamefont {S.}~\bibnamefont {Diehl}},\ }\bibfield  {title}
  {\bibinfo {title} {Ultracold quantum wires with localized losses: Many-body
  quantum zeno effect},\ }\href {https://doi.org/10.1103/PhysRevB.101.144301}
  {\bibfield  {journal} {\bibinfo  {journal} {Phys. Rev. B}\ }\textbf {\bibinfo
  {volume} {101}},\ \bibinfo {pages} {144301} (\bibinfo {year}
  {2020})}\BibitemShut {NoStop}%
\bibitem [{\citenamefont {Visuri}\ \emph {et~al.}(2023)\citenamefont {Visuri},
  \citenamefont {Giamarchi},\ and\ \citenamefont
  {Kollath}}]{visuri_losses_zeno_long}%
  \BibitemOpen
  \bibfield  {author} {\bibinfo {author} {\bibfnamefont {A.-M.}\ \bibnamefont
  {Visuri}}, \bibinfo {author} {\bibfnamefont {T.}~\bibnamefont {Giamarchi}},\
  and\ \bibinfo {author} {\bibfnamefont {C.}~\bibnamefont {Kollath}},\
  }\bibfield  {title} {\bibinfo {title} {Nonlinear transport in the presence of
  a local dissipation},\ }\href
  {https://doi.org/10.1103/PhysRevResearch.5.013195} {\bibfield  {journal}
  {\bibinfo  {journal} {Phys. Rev. Res.}\ }\textbf {\bibinfo {volume} {5}},\
  \bibinfo {pages} {013195} (\bibinfo {year} {2023})}\BibitemShut {NoStop}%
\bibitem [{\citenamefont {Huang}\ \emph {et~al.}(2023)\citenamefont {Huang},
  \citenamefont {Mohan}, \citenamefont {Visuri}, \citenamefont {Fabritius},
  \citenamefont {Talebi}, \citenamefont {Wili}, \citenamefont {Uchino},
  \citenamefont {Giamarchi},\ and\ \citenamefont {Esslinger}}]{huang_MAR_cold}%
  \BibitemOpen
  \bibfield  {author} {\bibinfo {author} {\bibfnamefont {M.-Z.}\ \bibnamefont
  {Huang}}, \bibinfo {author} {\bibfnamefont {J.}~\bibnamefont {Mohan}},
  \bibinfo {author} {\bibfnamefont {A.-M.}\ \bibnamefont {Visuri}}, \bibinfo
  {author} {\bibfnamefont {P.}~\bibnamefont {Fabritius}}, \bibinfo {author}
  {\bibfnamefont {M.}~\bibnamefont {Talebi}}, \bibinfo {author} {\bibfnamefont
  {S.}~\bibnamefont {Wili}}, \bibinfo {author} {\bibfnamefont {S.}~\bibnamefont
  {Uchino}}, \bibinfo {author} {\bibfnamefont {T.}~\bibnamefont {Giamarchi}},\
  and\ \bibinfo {author} {\bibfnamefont {T.}~\bibnamefont {Esslinger}},\
  }\bibfield  {title} {\bibinfo {title} {Superfluid signatures in a dissipative
  quantum point contact},\ }\href
  {https://doi.org/10.1103/PhysRevLett.130.200404} {\bibfield  {journal}
  {\bibinfo  {journal} {Phys. Rev. Lett.}\ }\textbf {\bibinfo {volume} {130}},\
  \bibinfo {pages} {200404} (\bibinfo {year} {2023})}\BibitemShut {NoStop}%
\bibitem [{\citenamefont {Bohn}\ and\ \citenamefont
  {Julienne}(1997)}]{Bohn_PhysRevA.56.1486}%
  \BibitemOpen
  \bibfield  {author} {\bibinfo {author} {\bibfnamefont {J.~L.}\ \bibnamefont
  {Bohn}}\ and\ \bibinfo {author} {\bibfnamefont {P.~S.}\ \bibnamefont
  {Julienne}},\ }\bibfield  {title} {\bibinfo {title} {Prospects for
  influencing scattering lengths with far-off-resonant light},\ }\href
  {https://doi.org/10.1103/PhysRevA.56.1486} {\bibfield  {journal} {\bibinfo
  {journal} {Phys. Rev. A}\ }\textbf {\bibinfo {volume} {56}},\ \bibinfo
  {pages} {1486} (\bibinfo {year} {1997})}\BibitemShut {NoStop}%
\bibitem [{\citenamefont {Cheianov}\ \emph {et~al.}(2006)\citenamefont
  {Cheianov}, \citenamefont {Smith},\ and\ \citenamefont
  {Zvonarev}}]{cheianov_threebody_losses}%
  \BibitemOpen
  \bibfield  {author} {\bibinfo {author} {\bibfnamefont {V.~V.}\ \bibnamefont
  {Cheianov}}, \bibinfo {author} {\bibfnamefont {H.}~\bibnamefont {Smith}},\
  and\ \bibinfo {author} {\bibfnamefont {M.~B.}\ \bibnamefont {Zvonarev}},\
  }\bibfield  {title} {\bibinfo {title} {Exact results for three-body
  correlations in a degenerate one-dimensional bose gas},\ }\href
  {https://doi.org/10.1103/PhysRevA.73.051604} {\bibfield  {journal} {\bibinfo
  {journal} {Phys. Rev. A}\ }\textbf {\bibinfo {volume} {73}},\ \bibinfo
  {pages} {051604} (\bibinfo {year} {2006})}\BibitemShut {NoStop}%
\bibitem [{\citenamefont {Braaten}\ \emph {et~al.}(2017)\citenamefont
  {Braaten}, \citenamefont {Hammer},\ and\ \citenamefont
  {Lepage}}]{Braaten_2017}%
  \BibitemOpen
  \bibfield  {author} {\bibinfo {author} {\bibfnamefont {E.}~\bibnamefont
  {Braaten}}, \bibinfo {author} {\bibfnamefont {H.-W.}\ \bibnamefont
  {Hammer}},\ and\ \bibinfo {author} {\bibfnamefont {G.~P.}\ \bibnamefont
  {Lepage}},\ }\bibfield  {title} {\bibinfo {title} {Lindblad equation for the
  inelastic loss of ultracold atoms},\ }\href
  {https://doi.org/10.1103/PhysRevA.95.012708} {\bibfield  {journal} {\bibinfo
  {journal} {Phys. Rev. A}\ }\textbf {\bibinfo {volume} {95}},\ \bibinfo
  {pages} {012708} (\bibinfo {year} {2017})}\BibitemShut {NoStop}%
\bibitem [{\citenamefont {Cazalilla}\ \emph {et~al.}(2011)\citenamefont
  {Cazalilla}, \citenamefont {Citro}, \citenamefont {Giamarchi}, \citenamefont
  {Orignac},\ and\ \citenamefont {Rigol}}]{RevModPhys.83.1405}%
  \BibitemOpen
  \bibfield  {author} {\bibinfo {author} {\bibfnamefont {M.~A.}\ \bibnamefont
  {Cazalilla}}, \bibinfo {author} {\bibfnamefont {R.}~\bibnamefont {Citro}},
  \bibinfo {author} {\bibfnamefont {T.}~\bibnamefont {Giamarchi}}, \bibinfo
  {author} {\bibfnamefont {E.}~\bibnamefont {Orignac}},\ and\ \bibinfo {author}
  {\bibfnamefont {M.}~\bibnamefont {Rigol}},\ }\bibfield  {title} {\bibinfo
  {title} {One dimensional bosons: From condensed matter systems to ultracold
  gases},\ }\href {https://doi.org/10.1103/RevModPhys.83.1405} {\bibfield
  {journal} {\bibinfo  {journal} {Rev. Mod. Phys.}\ }\textbf {\bibinfo {volume}
  {83}},\ \bibinfo {pages} {1405} (\bibinfo {year} {2011})}\BibitemShut
  {NoStop}%
\bibitem [{\citenamefont {Tomita}\ \emph {et~al.}(2019)\citenamefont {Tomita},
  \citenamefont {Nakajima}, \citenamefont {Takasu},\ and\ \citenamefont
  {Takahashi}}]{Tomita_PhysRevA_2019}%
  \BibitemOpen
  \bibfield  {author} {\bibinfo {author} {\bibfnamefont {T.}~\bibnamefont
  {Tomita}}, \bibinfo {author} {\bibfnamefont {S.}~\bibnamefont {Nakajima}},
  \bibinfo {author} {\bibfnamefont {Y.}~\bibnamefont {Takasu}},\ and\ \bibinfo
  {author} {\bibfnamefont {Y.}~\bibnamefont {Takahashi}},\ }\bibfield  {title}
  {\bibinfo {title} {Dissipative bose-hubbard system with intrinsic two-body
  loss},\ }\href {https://doi.org/10.1103/PhysRevA.99.031601} {\bibfield
  {journal} {\bibinfo  {journal} {Phys. Rev. A}\ }\textbf {\bibinfo {volume}
  {99}},\ \bibinfo {pages} {031601} (\bibinfo {year} {2019})}\BibitemShut
  {NoStop}%
\bibitem [{\citenamefont {Syassen}\ \emph
  {et~al.}(2008{\natexlab{b}})\citenamefont {Syassen}, \citenamefont {Bauer},
  \citenamefont {Lettner}, \citenamefont {Volz}, \citenamefont {Dietze},
  \citenamefont {Garc{\'\i}a-Ripoll}, \citenamefont {Cirac}, \citenamefont
  {Rempe},\ and\ \citenamefont {D{\"u}rr}}]{Syassen_science_2008}%
  \BibitemOpen
  \bibfield  {author} {\bibinfo {author} {\bibfnamefont {N.}~\bibnamefont
  {Syassen}}, \bibinfo {author} {\bibfnamefont {D.~M.}\ \bibnamefont {Bauer}},
  \bibinfo {author} {\bibfnamefont {M.}~\bibnamefont {Lettner}}, \bibinfo
  {author} {\bibfnamefont {T.}~\bibnamefont {Volz}}, \bibinfo {author}
  {\bibfnamefont {D.}~\bibnamefont {Dietze}}, \bibinfo {author} {\bibfnamefont
  {J.~J.}\ \bibnamefont {Garc{\'\i}a-Ripoll}}, \bibinfo {author} {\bibfnamefont
  {J.~I.}\ \bibnamefont {Cirac}}, \bibinfo {author} {\bibfnamefont
  {G.}~\bibnamefont {Rempe}},\ and\ \bibinfo {author} {\bibfnamefont
  {S.}~\bibnamefont {D{\"u}rr}},\ }\bibfield  {title} {\bibinfo {title} {Strong
  dissipation inhibits losses and induces correlations in cold molecular
  gases},\ }\href {https://doi.org/10.1126/science.1155309} {\bibfield
  {journal} {\bibinfo  {journal} {Science}\ }\textbf {\bibinfo {volume}
  {320}},\ \bibinfo {pages} {1329} (\bibinfo {year} {2008}{\natexlab{b}})},\
  \Eprint
  {https://arxiv.org/abs/https://www.science.org/doi/pdf/10.1126/science.1155309}
  {https://www.science.org/doi/pdf/10.1126/science.1155309} \BibitemShut
  {NoStop}%
\bibitem [{\citenamefont {García-Ripoll}\ \emph {et~al.}(2009)\citenamefont
  {García-Ripoll}, \citenamefont {Dürr}, \citenamefont {Syassen},
  \citenamefont {Bauer}, \citenamefont {Lettner}, \citenamefont {Rempe},\ and\
  \citenamefont {Cirac}}]{Garcia-Ripoll_2009}%
  \BibitemOpen
  \bibfield  {author} {\bibinfo {author} {\bibfnamefont {J.~J.}\ \bibnamefont
  {García-Ripoll}}, \bibinfo {author} {\bibfnamefont {S.}~\bibnamefont
  {Dürr}}, \bibinfo {author} {\bibfnamefont {N.}~\bibnamefont {Syassen}},
  \bibinfo {author} {\bibfnamefont {D.~M.}\ \bibnamefont {Bauer}}, \bibinfo
  {author} {\bibfnamefont {M.}~\bibnamefont {Lettner}}, \bibinfo {author}
  {\bibfnamefont {G.}~\bibnamefont {Rempe}},\ and\ \bibinfo {author}
  {\bibfnamefont {J.~I.}\ \bibnamefont {Cirac}},\ }\bibfield  {title} {\bibinfo
  {title} {Dissipation-induced hard-core boson gas in an optical lattice},\
  }\href {https://doi.org/10.1088/1367-2630/11/1/013053} {\bibfield  {journal}
  {\bibinfo  {journal} {New Journal of Physics}\ }\textbf {\bibinfo {volume}
  {11}},\ \bibinfo {pages} {013053} (\bibinfo {year} {2009})}\BibitemShut
  {NoStop}%
\bibitem [{\citenamefont {Cazalilla}(2003)}]{Cazalilla_PhysRevA.67.053606}%
  \BibitemOpen
  \bibfield  {author} {\bibinfo {author} {\bibfnamefont {M.~A.}\ \bibnamefont
  {Cazalilla}},\ }\bibfield  {title} {\bibinfo {title} {One-dimensional optical
  lattices and impenetrable bosons},\ }\href
  {https://doi.org/10.1103/PhysRevA.67.053606} {\bibfield  {journal} {\bibinfo
  {journal} {Phys. Rev. A}\ }\textbf {\bibinfo {volume} {67}},\ \bibinfo
  {pages} {053606} (\bibinfo {year} {2003})}\BibitemShut {NoStop}%
\bibitem [{\citenamefont {Rossini}\ \emph {et~al.}(2021)\citenamefont
  {Rossini}, \citenamefont {Ghermaoui}, \citenamefont {Aguilera}, \citenamefont
  {Vatr\'e}, \citenamefont {Bouganne}, \citenamefont {Beugnon}, \citenamefont
  {Gerbier},\ and\ \citenamefont {Mazza}}]{Rossini_PhysRevA.103.L060201}%
  \BibitemOpen
  \bibfield  {author} {\bibinfo {author} {\bibfnamefont {D.}~\bibnamefont
  {Rossini}}, \bibinfo {author} {\bibfnamefont {A.}~\bibnamefont {Ghermaoui}},
  \bibinfo {author} {\bibfnamefont {M.~B.}\ \bibnamefont {Aguilera}}, \bibinfo
  {author} {\bibfnamefont {R.}~\bibnamefont {Vatr\'e}}, \bibinfo {author}
  {\bibfnamefont {R.}~\bibnamefont {Bouganne}}, \bibinfo {author}
  {\bibfnamefont {J.}~\bibnamefont {Beugnon}}, \bibinfo {author} {\bibfnamefont
  {F.}~\bibnamefont {Gerbier}},\ and\ \bibinfo {author} {\bibfnamefont
  {L.}~\bibnamefont {Mazza}},\ }\bibfield  {title} {\bibinfo {title} {Strong
  correlations in lossy one-dimensional quantum gases: From the quantum zeno
  effect to the generalized gibbs ensemble},\ }\href
  {https://doi.org/10.1103/PhysRevA.103.L060201} {\bibfield  {journal}
  {\bibinfo  {journal} {Phys. Rev. A}\ }\textbf {\bibinfo {volume} {103}},\
  \bibinfo {pages} {L060201} (\bibinfo {year} {2021})}\BibitemShut {NoStop}%
\bibitem [{\citenamefont {Perfetto}\ \emph {et~al.}(2022)\citenamefont
  {Perfetto}, \citenamefont {Carollo}, \citenamefont {Garrahan},\ and\
  \citenamefont {Lesanovsky}}]{perfetto2022reactionlimited}%
  \BibitemOpen
  \bibfield  {author} {\bibinfo {author} {\bibfnamefont {G.}~\bibnamefont
  {Perfetto}}, \bibinfo {author} {\bibfnamefont {F.}~\bibnamefont {Carollo}},
  \bibinfo {author} {\bibfnamefont {J.~P.}\ \bibnamefont {Garrahan}},\ and\
  \bibinfo {author} {\bibfnamefont {I.}~\bibnamefont {Lesanovsky}},\
  }\href@noop {} {\bibinfo {title} {Reaction-limited quantum reaction-diffusion
  dynamics}} (\bibinfo {year} {2022}),\ \Eprint
  {https://arxiv.org/abs/2209.09784} {arXiv:2209.09784 [cond-mat.stat-mech]}
  \BibitemShut {NoStop}%
\bibitem [{\citenamefont {Cazalilla}\ and\ \citenamefont
  {Rey}(2014)}]{Cazalilla_2014}%
  \BibitemOpen
  \bibfield  {author} {\bibinfo {author} {\bibfnamefont {M.~A.}\ \bibnamefont
  {Cazalilla}}\ and\ \bibinfo {author} {\bibfnamefont {A.~M.}\ \bibnamefont
  {Rey}},\ }\bibfield  {title} {\bibinfo {title} {Ultracold fermi gases with
  emergent su(n) symmetry},\ }\href
  {https://doi.org/10.1088/0034-4885/77/12/124401} {\bibfield  {journal}
  {\bibinfo  {journal} {Reports on Progress in Physics}\ }\textbf {\bibinfo
  {volume} {77}},\ \bibinfo {pages} {124401} (\bibinfo {year}
  {2014})}\BibitemShut {NoStop}%
\bibitem [{\citenamefont {Cazalilla}\ \emph {et~al.}(2009)\citenamefont
  {Cazalilla}, \citenamefont {Ho},\ and\ \citenamefont
  {Ueda}}]{Cazalilla_2009}%
  \BibitemOpen
  \bibfield  {author} {\bibinfo {author} {\bibfnamefont {M.~A.}\ \bibnamefont
  {Cazalilla}}, \bibinfo {author} {\bibfnamefont {A.~F.}\ \bibnamefont {Ho}},\
  and\ \bibinfo {author} {\bibfnamefont {M.}~\bibnamefont {Ueda}},\ }\bibfield
  {title} {\bibinfo {title} {Ultracold gases of ytterbium: ferromagnetism and
  mott states in an su(6) fermi system},\ }\href
  {https://doi.org/10.1088/1367-2630/11/10/103033} {\bibfield  {journal}
  {\bibinfo  {journal} {New Journal of Physics}\ }\textbf {\bibinfo {volume}
  {11}},\ \bibinfo {pages} {103033} (\bibinfo {year} {2009})}\BibitemShut
  {NoStop}%
\bibitem [{\citenamefont {Theis}\ \emph {et~al.}(2004)\citenamefont {Theis},
  \citenamefont {Thalhammer}, \citenamefont {Winkler}, \citenamefont {Hellwig},
  \citenamefont {Ruff}, \citenamefont {Grimm},\ and\ \citenamefont
  {Denschlag}}]{Thies_PhysRevLett.93.123001}%
  \BibitemOpen
  \bibfield  {author} {\bibinfo {author} {\bibfnamefont {M.}~\bibnamefont
  {Theis}}, \bibinfo {author} {\bibfnamefont {G.}~\bibnamefont {Thalhammer}},
  \bibinfo {author} {\bibfnamefont {K.}~\bibnamefont {Winkler}}, \bibinfo
  {author} {\bibfnamefont {M.}~\bibnamefont {Hellwig}}, \bibinfo {author}
  {\bibfnamefont {G.}~\bibnamefont {Ruff}}, \bibinfo {author} {\bibfnamefont
  {R.}~\bibnamefont {Grimm}},\ and\ \bibinfo {author} {\bibfnamefont {J.~H.}\
  \bibnamefont {Denschlag}},\ }\bibfield  {title} {\bibinfo {title} {Tuning the
  scattering length with an optically induced feshbach resonance},\ }\href
  {https://doi.org/10.1103/PhysRevLett.93.123001} {\bibfield  {journal}
  {\bibinfo  {journal} {Phys. Rev. Lett.}\ }\textbf {\bibinfo {volume} {93}},\
  \bibinfo {pages} {123001} (\bibinfo {year} {2004})}\BibitemShut {NoStop}%
\bibitem [{\citenamefont {Ciury\l{}o}\ \emph {et~al.}(2004)\citenamefont
  {Ciury\l{}o}, \citenamefont {Tiesinga}, \citenamefont {Kotochigova},\ and\
  \citenamefont {Julienne}}]{Ciurylo_PhysRevA_2004}%
  \BibitemOpen
  \bibfield  {author} {\bibinfo {author} {\bibfnamefont {R.}~\bibnamefont
  {Ciury\l{}o}}, \bibinfo {author} {\bibfnamefont {E.}~\bibnamefont
  {Tiesinga}}, \bibinfo {author} {\bibfnamefont {S.}~\bibnamefont
  {Kotochigova}},\ and\ \bibinfo {author} {\bibfnamefont {P.~S.}\ \bibnamefont
  {Julienne}},\ }\bibfield  {title} {\bibinfo {title} {Photoassociation
  spectroscopy of cold alkaline-earth-metal atoms near the intercombination
  line},\ }\href {https://doi.org/10.1103/PhysRevA.70.062710} {\bibfield
  {journal} {\bibinfo  {journal} {Phys. Rev. A}\ }\textbf {\bibinfo {volume}
  {70}},\ \bibinfo {pages} {062710} (\bibinfo {year} {2004})}\BibitemShut
  {NoStop}%
\bibitem [{\citenamefont {Enomoto}\ \emph {et~al.}(2008)\citenamefont
  {Enomoto}, \citenamefont {Kasa}, \citenamefont {Kitagawa},\ and\
  \citenamefont {Takahashi}}]{enomoto_PhysRevLett.101.203201}%
  \BibitemOpen
  \bibfield  {author} {\bibinfo {author} {\bibfnamefont {K.}~\bibnamefont
  {Enomoto}}, \bibinfo {author} {\bibfnamefont {K.}~\bibnamefont {Kasa}},
  \bibinfo {author} {\bibfnamefont {M.}~\bibnamefont {Kitagawa}},\ and\
  \bibinfo {author} {\bibfnamefont {Y.}~\bibnamefont {Takahashi}},\ }\bibfield
  {title} {\bibinfo {title} {Optical feshbach resonance using the
  intercombination transition},\ }\href
  {https://doi.org/10.1103/PhysRevLett.101.203201} {\bibfield  {journal}
  {\bibinfo  {journal} {Phys. Rev. Lett.}\ }\textbf {\bibinfo {volume} {101}},\
  \bibinfo {pages} {203201} (\bibinfo {year} {2008})}\BibitemShut {NoStop}%
\bibitem [{\citenamefont {Napolitano}\ \emph {et~al.}(1994)\citenamefont
  {Napolitano}, \citenamefont {Weiner}, \citenamefont {Williams},\ and\
  \citenamefont {Julienne}}]{Napolitano_PhysRevLett.73.1352}%
  \BibitemOpen
  \bibfield  {author} {\bibinfo {author} {\bibfnamefont {R.}~\bibnamefont
  {Napolitano}}, \bibinfo {author} {\bibfnamefont {J.}~\bibnamefont {Weiner}},
  \bibinfo {author} {\bibfnamefont {C.~J.}\ \bibnamefont {Williams}},\ and\
  \bibinfo {author} {\bibfnamefont {P.~S.}\ \bibnamefont {Julienne}},\
  }\bibfield  {title} {\bibinfo {title} {Line shapes of high resolution
  photoassociation spectra of optically cooled atoms},\ }\href
  {https://doi.org/10.1103/PhysRevLett.73.1352} {\bibfield  {journal} {\bibinfo
   {journal} {Phys. Rev. Lett.}\ }\textbf {\bibinfo {volume} {73}},\ \bibinfo
  {pages} {1352} (\bibinfo {year} {1994})}\BibitemShut {NoStop}%
\bibitem [{\citenamefont {Yamazaki}\ \emph {et~al.}(2013)\citenamefont
  {Yamazaki}, \citenamefont {Taie}, \citenamefont {Sugawa}, \citenamefont
  {Enomoto},\ and\ \citenamefont {Takahashi}}]{Yamazaki_PhysRevA.87.010704}%
  \BibitemOpen
  \bibfield  {author} {\bibinfo {author} {\bibfnamefont {R.}~\bibnamefont
  {Yamazaki}}, \bibinfo {author} {\bibfnamefont {S.}~\bibnamefont {Taie}},
  \bibinfo {author} {\bibfnamefont {S.}~\bibnamefont {Sugawa}}, \bibinfo
  {author} {\bibfnamefont {K.}~\bibnamefont {Enomoto}},\ and\ \bibinfo {author}
  {\bibfnamefont {Y.}~\bibnamefont {Takahashi}},\ }\bibfield  {title} {\bibinfo
  {title} {Observation of a $p$-wave optical feshbach resonance},\ }\href
  {https://doi.org/10.1103/PhysRevA.87.010704} {\bibfield  {journal} {\bibinfo
  {journal} {Phys. Rev. A}\ }\textbf {\bibinfo {volume} {87}},\ \bibinfo
  {pages} {010704} (\bibinfo {year} {2013})}\BibitemShut {NoStop}%
\bibitem [{\citenamefont {Kim}\ \emph {et~al.}(2016)\citenamefont {Kim},
  \citenamefont {Lee}, \citenamefont {Lee}, \citenamefont {Shin},\ and\
  \citenamefont {Mun}}]{Kim_PhysRevA.94.042703}%
  \BibitemOpen
  \bibfield  {author} {\bibinfo {author} {\bibfnamefont {M.-S.}\ \bibnamefont
  {Kim}}, \bibinfo {author} {\bibfnamefont {J.}~\bibnamefont {Lee}}, \bibinfo
  {author} {\bibfnamefont {J.~H.}\ \bibnamefont {Lee}}, \bibinfo {author}
  {\bibfnamefont {Y.}~\bibnamefont {Shin}},\ and\ \bibinfo {author}
  {\bibfnamefont {J.}~\bibnamefont {Mun}},\ }\bibfield  {title} {\bibinfo
  {title} {Measurements of optical feshbach resonances of $^{174}\mathrm{Yb}$
  atoms},\ }\href {https://doi.org/10.1103/PhysRevA.94.042703} {\bibfield
  {journal} {\bibinfo  {journal} {Phys. Rev. A}\ }\textbf {\bibinfo {volume}
  {94}},\ \bibinfo {pages} {042703} (\bibinfo {year} {2016})}\BibitemShut
  {NoStop}%
\bibitem [{\citenamefont {Chin}\ \emph {et~al.}(2010)\citenamefont {Chin},
  \citenamefont {Grimm}, \citenamefont {Julienne},\ and\ \citenamefont
  {Tiesinga}}]{RevModPhys.82.1225}%
  \BibitemOpen
  \bibfield  {author} {\bibinfo {author} {\bibfnamefont {C.}~\bibnamefont
  {Chin}}, \bibinfo {author} {\bibfnamefont {R.}~\bibnamefont {Grimm}},
  \bibinfo {author} {\bibfnamefont {P.}~\bibnamefont {Julienne}},\ and\
  \bibinfo {author} {\bibfnamefont {E.}~\bibnamefont {Tiesinga}},\ }\bibfield
  {title} {\bibinfo {title} {Feshbach resonances in ultracold gases},\ }\href
  {https://doi.org/10.1103/RevModPhys.82.1225} {\bibfield  {journal} {\bibinfo
  {journal} {Rev. Mod. Phys.}\ }\textbf {\bibinfo {volume} {82}},\ \bibinfo
  {pages} {1225} (\bibinfo {year} {2010})}\BibitemShut {NoStop}%
\bibitem [{\citenamefont {Blatt}\ \emph {et~al.}(2011)\citenamefont {Blatt},
  \citenamefont {Nicholson}, \citenamefont {Bloom}, \citenamefont {Williams},
  \citenamefont {Thomsen}, \citenamefont {Julienne},\ and\ \citenamefont
  {Ye}}]{Blatt_PhysRevLett_2011}%
  \BibitemOpen
  \bibfield  {author} {\bibinfo {author} {\bibfnamefont {S.}~\bibnamefont
  {Blatt}}, \bibinfo {author} {\bibfnamefont {T.~L.}\ \bibnamefont
  {Nicholson}}, \bibinfo {author} {\bibfnamefont {B.~J.}\ \bibnamefont
  {Bloom}}, \bibinfo {author} {\bibfnamefont {J.~R.}\ \bibnamefont {Williams}},
  \bibinfo {author} {\bibfnamefont {J.~W.}\ \bibnamefont {Thomsen}}, \bibinfo
  {author} {\bibfnamefont {P.~S.}\ \bibnamefont {Julienne}},\ and\ \bibinfo
  {author} {\bibfnamefont {J.}~\bibnamefont {Ye}},\ }\bibfield  {title}
  {\bibinfo {title} {Measurement of optical feshbach resonances in an ideal
  gas},\ }\href {https://doi.org/10.1103/PhysRevLett.107.073202} {\bibfield
  {journal} {\bibinfo  {journal} {Phys. Rev. Lett.}\ }\textbf {\bibinfo
  {volume} {107}},\ \bibinfo {pages} {073202} (\bibinfo {year}
  {2011})}\BibitemShut {NoStop}%
\bibitem [{\citenamefont {Giamarchi}(2003)}]{Giamarchi_1dbook}%
  \BibitemOpen
  \bibfield  {author} {\bibinfo {author} {\bibfnamefont {T.}~\bibnamefont
  {Giamarchi}},\ }\href
  {https://doi.org/10.1093/acprof:oso/9780198525004.001.0001} {\emph {\bibinfo
  {title} {{Quantum Physics in One Dimension}}}}\ (\bibinfo  {publisher}
  {Oxford University Press},\ \bibinfo {year} {2003})\BibitemShut {NoStop}%
\bibitem [{\citenamefont {Cazalilla}(2004)}]{Cazalilla_2004}%
  \BibitemOpen
  \bibfield  {author} {\bibinfo {author} {\bibfnamefont {M.~A.}\ \bibnamefont
  {Cazalilla}},\ }\bibfield  {title} {\bibinfo {title} {Bosonizing
  one-dimensional cold atomic gases},\ }\href
  {https://doi.org/10.1088/0953-4075/37/7/051} {\bibfield  {journal} {\bibinfo
  {journal} {Journal of Physics B: Atomic, Molecular and Optical Physics}\
  }\textbf {\bibinfo {volume} {37}},\ \bibinfo {pages} {S1} (\bibinfo {year}
  {2004})}\BibitemShut {NoStop}%
\bibitem [{\citenamefont {Sieberer}\ \emph
  {et~al.}(2016{\natexlab{b}})\citenamefont {Sieberer}, \citenamefont
  {Buchhold},\ and\ \citenamefont {Diehl}}]{Sieberer2016}%
  \BibitemOpen
  \bibfield  {author} {\bibinfo {author} {\bibfnamefont {L.~M.}\ \bibnamefont
  {Sieberer}}, \bibinfo {author} {\bibfnamefont {M.}~\bibnamefont {Buchhold}},\
  and\ \bibinfo {author} {\bibfnamefont {S.}~\bibnamefont {Diehl}},\ }\bibfield
   {title} {\bibinfo {title} {Keldysh field theory for driven open quantum
  systems},\ }\href {https://doi.org/10.1088/0034-4885/79/9/096001} {\bibfield
  {journal} {\bibinfo  {journal} {Reports on Progress in Physics}\ }\textbf
  {\bibinfo {volume} {79}},\ \bibinfo {pages} {096001} (\bibinfo {year}
  {2016}{\natexlab{b}})}\BibitemShut {NoStop}%
\bibitem [{\citenamefont {Tonielli}(2016)}]{TONIELLI_2016}%
  \BibitemOpen
  \bibfield  {author} {\bibinfo {author} {\bibfnamefont {F.}~\bibnamefont
  {Tonielli}},\ }\bibfield  {title} {\bibinfo {title} {Keldysh field theory for
  dissipation-induced states of fermions},\ }\href
  {https://etd.adm.unipi.it/t/etd-05052016-191618/} {\bibfield  {journal}
  {\bibinfo  {journal} {Phd thesis, University of Pisa}\ } (\bibinfo {year}
  {2016})}\BibitemShut {NoStop}%
\bibitem [{\citenamefont {Weiss}(2008)}]{Weissbook2008}%
  \BibitemOpen
  \bibfield  {author} {\bibinfo {author} {\bibfnamefont {U.}~\bibnamefont
  {Weiss}},\ }\href@noop {} {\emph {\bibinfo {title} {Quantum Dissipative
  Systems}}}\ (\bibinfo  {publisher} {World Scientific, Series in Modern
  Condensed Matter Vol.~13},\ \bibinfo {address} {Singapore},\ \bibinfo {year}
  {2008})\BibitemShut {NoStop}%
\end{thebibliography}%

\end{document}